\documentclass[twocolumn,iop]{emulateapj}

\usepackage{graphics}
\graphicspath{{Figures/}{./}}

\newcommand{\kms}{$\rm km\,s^{-1}$}
\newcommand{\code}[1]{\texttt{#1}}

\begin{document}

\title{The Ninth Data Release of the Sloan Digital Sky Survey: 
First Spectroscopic Data from the \mbox{SDSS-III} Baryon Oscillation Spectroscopic Survey}

\author{
Christopher~P.~Ahn\altaffilmark{1},    
Rachael~Alexandroff\altaffilmark{2},
Carlos~{Allende~Prieto}\altaffilmark{3,4},
 Scott~F.~Anderson\altaffilmark{5},
Timothy~Anderton\altaffilmark{1},
Brett~H.~Andrews\altaffilmark{6},
\'Eric~Aubourg\altaffilmark{7}
Stephen~Bailey\altaffilmark{8},
Rory~Barnes\altaffilmark{5},
Julian~Bautista\altaffilmark{7}
% ...
}

\altaffiltext{1}{
Department of Physics and Astronomy, 
University of Utah, Salt Lake City, UT 84112, USA.
}

\altaffiltext{2}{
Department of Astrophysical Sciences, Princeton University, 
Princeton, NJ 08544, USA.
}

\altaffiltext{3}{
Instituto de Astrof{\'\i}sica de Canarias (IAC), C/V{\'\i}a L\'actea,
s/n, E-38200, La Laguna, Tenerife, Spain.
}

\altaffiltext{4}{
Departamento de Astrof\'{\i}sica, 
Universidad de La Laguna, 
E-38206, La Laguna, Tenerife, Spain.
}

\altaffiltext{5}{
Department of Astronomy, University of Washington, 
Box 351580, Seattle, WA 98195, USA.
}

\altaffiltext{6}{
Department of Astronomy, 
Ohio State University, 140 West 18th Avenue, Columbus, OH 43210, USA.
}

\altaffiltext{7}{
APC, University of Paris Diderot, CNRS/IN2P3, CEA/IRFU, Observatoire de Paris, Sorbonne Paris Cit\'e, France.
}

\altaffiltext{8}{
Lawrence Berkeley National Laboratory, One Cyclotron Road,
Berkeley, CA 94720, USA.
}

% ...

\shorttitle{SDSS DR9}

\begin{abstract}

The Sloan Digital Sky Survey III (\mbox{SDSS-III}) presents the first
spectroscopic data from the Baryon Oscillation Spectroscopic Survey
(BOSS). 
This ninth data release (DR9) of the SDSS project includes
535,995 new galaxy spectra (median $z\sim 0.52$),
102,100 new quasar spectra (median $z\sim 2.32$), 
and 90,897 new stellar spectra,  
along with the data presented in previous data releases.
These spectra were obtained with the new BOSS spectrograph and were
taken between 2009 December and 2011 July.  
In addition, the stellar parameters pipeline, which determines radial
velocities, surface temperatures, surface gravities, and
metallicities of stars, has been updated
and refined with improvements in temperature estimates for stars
with $T_{\rm eff}<5000$~K and in metallicity estimates for stars with $\rm [Fe/H]>-0.5$.   
DR9 includes new stellar
parameters for all stars presented in DR8,
including stars from SDSS-I and II, as well as those observed
as part of the 
\mbox{SDSS-III}
Sloan Extension for
Galactic Understanding and Exploration-2 (SEGUE-2). 

The astrometry error introduced in the DR8 imaging catalogs has been corrected in the DR9 data products.
The next data release for \mbox{SDSS-III} will be in Summer 2013, which will
present the first data from the Apache Point Observatory Galactic
Evolution Experiment (APOGEE) along with another year
of data from BOSS, followed by the final \mbox{SDSS-III} data release in December
2014. 
\end{abstract}

\section{Introduction}
\label{sec:introduction}

The Sloan Digital Sky Survey III (\mbox{SDSS-III}; \citealt{eisenstein11}) is an extension of the
SDSS-I and II projects \citep{york00}.  It uses the dedicated 2.5-meter
wide-field Sloan Foundation Telescope \citep{gunn06} at Apache Point Observatory (APO),
and fiber-fed multi-object spectrographs to carry out four surveys
to study dark energy through observations of
distant galaxies and quasars 
(the Baryon Oscillation Sky Survey; BOSS), 
to understand the structure of the Milky Way Galaxy 
(the Sloan Extension for Galaxy Understanding and Exploration; SEGUE-2, 
and the APO Galactic Evolution Experiment; APOGEE), 
and to search for
extrasolar planets (the Multi-object APO Radial Velocity Exoplanet
Large-area Survey; MARVELS).  \mbox{SDSS-III} commenced in Fall 2008, and
will carry out observations for six years through Summer 2014.  
The first data release of
this phase of SDSS (and the eighth release overall; DR8;
\citealt{DR8}) was made public in Winter 2011.  In addition to all the
data from SDSS-I and II \citep{DR7}, DR8 included additional five-band imaging
data over 2500 deg$^2$ over the Southern Galactic Cap, as well as
stellar spectra from SEGUE-2.  

This paper presents the ninth data release (DR9) from SDSS, including all
survey-quality data from BOSS gathered through 2011 July.  BOSS
\citep{dawson12} uses new spectrographs \citep{smee12} to obtain
spectra of galaxies with $0.15 < z < 0.8$ and quasars with 
$2.15 < z < 3.5$ 
to measure the scale of the baryon oscillation peak in the
correlation function of matter in order to probe the geometry and
dynamics of the universe.  DR9 includes the first year of BOSS data,
and this paper describes the characteristics of these data (summarized in \S\ref{sec:scope}), with a
particular emphasis on how it differs from the spectroscopy carried
out in SDSS-I and SDSS-II (\S\ref{sec:BOSS}).  

The erratum to the DR8 paper \citep{DR8_erratum} describes a
systematic error in the astrometry in the imaging catalogs in DR8.
This has now been fixed, as we describe in \S\ref{sec:astrometry}. 

The SEGUE Stellar Parameters Pipeline (SSPP) fits detailed models to
the spectrum of each star, to determine surface temperatures,
metallicities, and gravities.  It has been continuously improved since
its introduction in the sixth data release (DR6, \citealt{DR6}; see
also \citealt{lee08a}).  In \S\ref{sec:SSPP}, we describe the
improvements since DR8 that are incorporated into the DR9 outputs. 

Section~\ref{sec:distribution} describes how one can access the DR9 data,
and we conclude and outline the planned future data releases in \S\ref{sec:conclusions}.

\section{Scope of DR9}
\label{sec:scope}

DR9 presents the release of the first 1.5 years of data from the
\mbox{SDSS-III} BOSS spectroscopic survey.  BOSS started commissioning in
early Fall 2009, and began survey-quality observations on the night of
2009 December 5 (UTC-7; MJD 55171).  
All processed data from that date until the
summer telescope shutdown\footnote[98]{The SDSS telescope pauses 
  science operations during the month-long ``monsoon'' in July/August in the
  southwestern United States.  This time is used for telescope
  maintenance and engineering work.} in 2011 July are included in DR9.  
All raw data taken by the BOSS spectrograph from the start of
commissioning (2009 September) through and including 2011 July 10 (MJD 55752) are also
available as flat files as part of the DR9 release, although the
commissioning data are of quite poor quality, and don't always include
data from both spectrographs.  DR9 also includes
the spectroscopic data from SDSS-I/II and SEGUE2; it is unchanged
since DR8.  

\begin{figure*}
\plotone{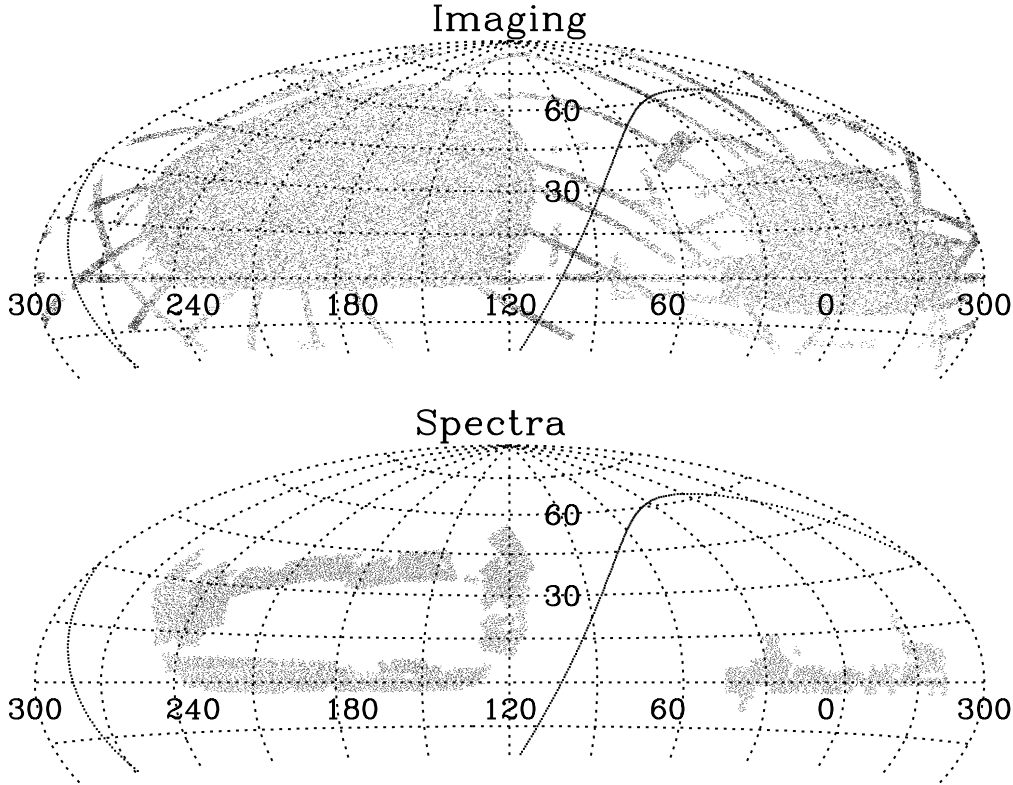}
\caption{The distribution on the sky of all SDSS imaging (top; same as
  DR8) and BOSS DR9 spectroscopy (bottom) in equatorial coordinates 
($\alpha=0^\circ$ is offset to the right in this projection).  
The Galactic equatorial plane is shown by the solid line.
To make
the image for BOSS spectroscopy, we simply plotted a sparse version of
the BOSS quasar catalog \citep{paris12}.}
\label{fig:skydist}
\end{figure*}

The details of the data included in DR9 are summarized in
Table~\ref{table:dr9_contents}, and the footprints of the imaging and
spectroscopic data are shown
in Figure~\ref{fig:skydist}.  The imaging data and imaging catalogs
are the same as in DR8, with the key update of an improved astrometric
solution to correct an error affecting objects at high
declinations~\citep{DR8_erratum}. 

\begin{figure}
\plotone{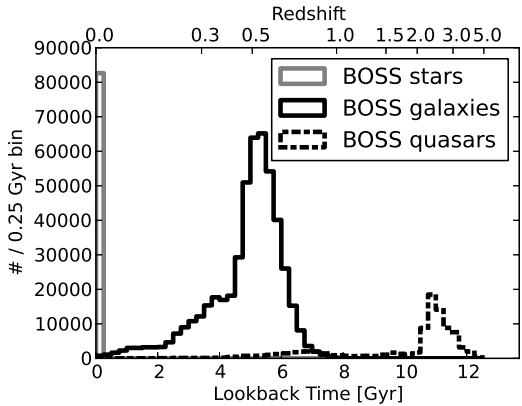}
\caption{The distribution with lookback time of the 
82,645 stars; 493,845 galaxies; and 
93,003 quasars with spectra in DR9 BOSS.
Lookback time is based on the observed redshift under the assumption of
a flat $\Lambda$CDM cosmology ($\Omega_M$,$\Omega_\Lambda$,$h$)=(0.272,0.728,0.71) 
consistent with the joint cosmological analysis of WMAP7~\citep{komatsu11}.
}
\label{fig:boss_star_galaxy_qso}
\end{figure}

\begin{figure}
\plotone{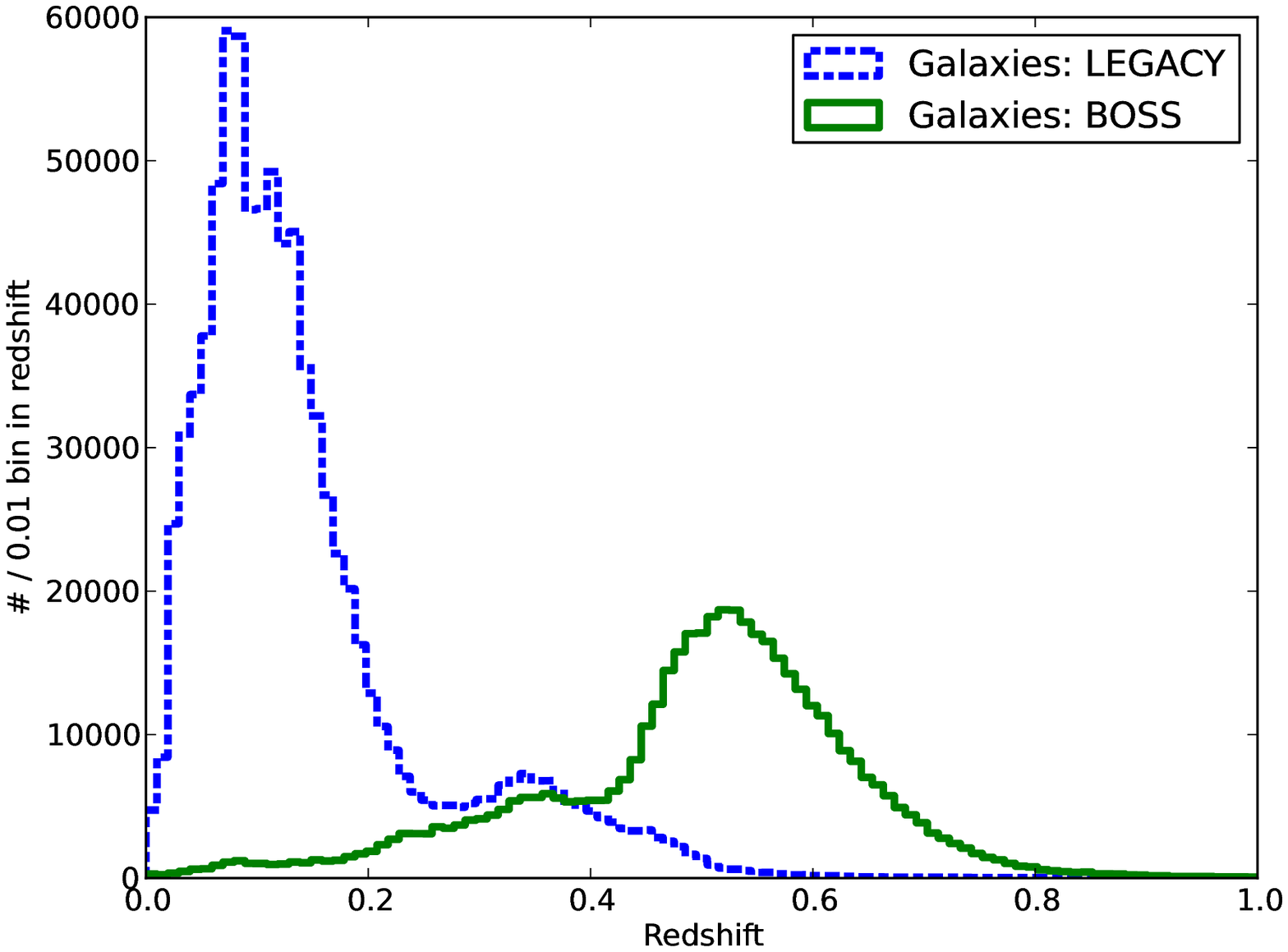}
\plotone{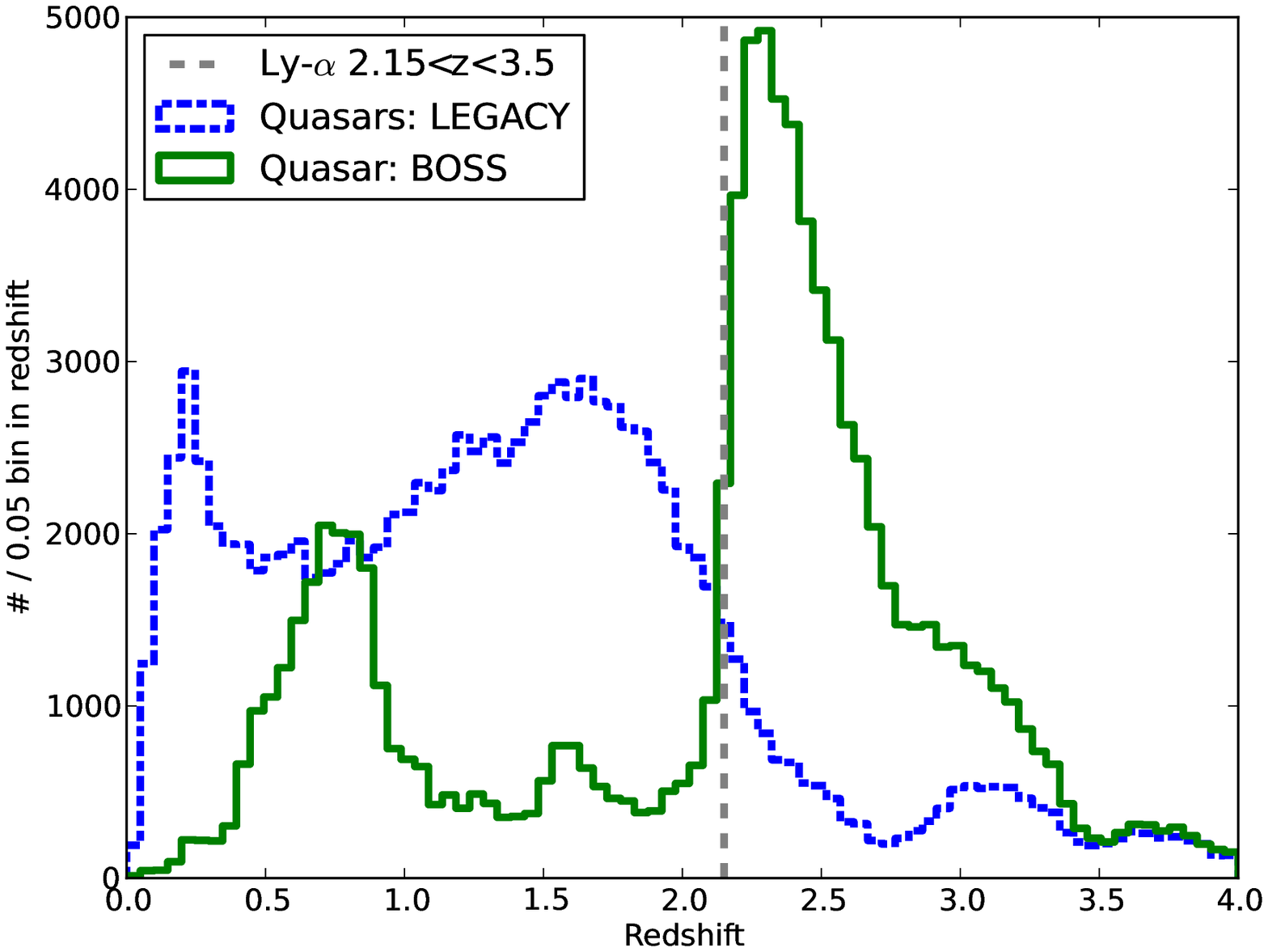}
\caption{$N(z)$ of BOSS spectra in DR9 compared to that of the
  SDSS-I/II Legacy spectra for galaxies (left) and quasars (right).
  BOSS' focus on galaxies with $0.4<z<0.6$ and quasars with $z >
  2.15$ is apparent.  The BOSS quasars at
  $0.5<z<0.9$ are selected because of a degeneracy in color space
  between these lower-redshift quasars and those at $z>2.15$. 
}
\label{fig:boss_galaxy_qso_vs_sdss}
\end{figure}

Fig.~\ref{fig:boss_star_galaxy_qso} presents the distribution with
look-back time of
spectroscopically confirmed stars, galaxies, and quasars 
 from BOSS in the DR9 data set.   
Fig.~\ref{fig:boss_galaxy_qso_vs_sdss} compares these distributions to
those of all previous SDSS spectra of galaxies and quasars.

All data released with DR9 are publicly available at
\url{http://www.sdss3.org/dr9}. 

\begin{deluxetable}{lrr}
\tablecaption{Contents of DR9\label{table:dr9_contents}}
\startdata
\cutinhead{Imaging\tablenotemark{a}}
& Total & Unique\tablenotemark{b} \\
 Area Imaged   & 31,637 deg$^2$ & 14,555 deg$^2$ \\
 Cataloged Objects & 1,231,051,050 & 469,053,874 \\   \cutinhead{New BOSS Spectroscopy\tablenotemark{c}}
& Total & Unique\tablenotemark{b} \\
Spectroscopic footprint effective area & \nodata & 3275~deg$^2$ \\
Plates\tablenotemark{d}             &     831 &     819 \\
Spectra observed\tablenotemark{e}   & 829,073 & 763,425 \\
 & & \\
Galaxies                            & 535,995 & 493,845 \\
CMASS galaxies                      & 336,695 & 309,307 \\
LOWZ galaxies                       & 110,427 & 102,890 \\
All Quasars                         & 102,100 &  93,003 \\
Main Quasars\tablenotemark{f}       &  85,977 &  79,570 \\
Main Quasars, $2.15<z<3.5$\tablenotemark{g} &  59,783 &  55,047 \\
Ancillary program spectra           &  32,381 &  28,968 \\
Stars                               &  90,897 &  82,645 \\   Standard stars                      &  16,905 &  14,915 \\
Sky spectra                         &  78,573 &  75,850 \\
\cutinhead{All Spectroscopy from \mbox{SDSS-I/II/III}} 
 Total number of spectra & 2,674,200 \\
 Total number of useful spectra\tablenotemark{h} & 2,598,033 \\  \quad  Galaxies   & 1,457,002  \\  \quad  Quasars    &   228,468  \\  \quad  Stars      &   668,054  \\  \quad  Sky        &   181,619  \\  \quad  Unclassified\tablenotemark{i} & 62,890   \enddata
\tablenotetext{a}{These numbers are unchanged since DR8.}
\tablenotetext{b}{Removing all duplicates and overlaps.}
\tablenotetext{c}{See \citet{bolton12} for full details.}
\tablenotetext{d}{Twelve plates of the 831 observed plates were re-plugged and re-observed for calibration purposes. Six of the 819 unique plates are different drillings of the same tiling objects.}
\tablenotetext{e}{This excludes the small fraction of the observations through broken fibers or those that fell out of their holes.  There were 831,000 spectra attempted.}
\tablenotetext{f}{This counts only quasars from the main survey
  (\S\ref{sec:quasars}), and does not include those from ancillary
  programs (\S\ref{sec:ancillary}) or that were used for calibration
  purposes.}  
\tablenotetext{g}{Quasars with redshifts in the range $2.15<z<3.5$
  provide the most signal in the BOSS spectra of the 
  Ly-$\alpha$ forest.} 
\tablenotetext{h}{Spectra on good or marginal plates.  ``Spectrum''
  refers to a combined set of sub-exposures that define a completed
  plate.  Duplicates are from plates that were observed more than
  once, or are objects that were observed on overlapping plates.}
\tablenotetext{i}{Non-sky spectra for which the automated
redshift/classification pipeline \citep{bolton12} gave unreliable
results, as indicated by the \code{ZWARNING} flag.}
\end{deluxetable}

\section{The Baryon Oscillation Spectroscopic Survey}
\label{sec:BOSS}

When the Universe was radiation-dominated, sound
waves propagated through the radiation-matter fluid at 
a significant fraction of 
the speed of light.  They slowed dramatically after
matter-radiation equality, and were frozen in after recombination.
Sound waves propagating from overdensities thus propagated
a given distance, roughly 150 comoving Mpc (given standard cosmological
parameters) from the initial perturbations; the resulting
overdensity gives an excess in the clustering of matter at this
scale.  This is the origin of the oscillations seen in the power
spectrum of the Cosmic Microwave Background (e.g.,
\citealt{komatsu11}), and was first conclusively seen in the
clustering of galaxies 
from 
the Two Degree Field Galaxy Redshift Survey \citep{cole05}
 and
the SDSS \citep{eisenstein05}. 
This feature in the galaxy or matter
correlation function or power spectrum is a {\em standard ruler};
measuring it as a function of redshift gives a powerful constraint on
cosmological models \citep[e.g.,][]{weinberg12}.  

The initial SDSS detection of the baryon oscillation feature
(\citealt{eisenstein05}; see also
\citealt{tegmark06,percival10,padmanabhan12a}) was based upon a galaxy sample
at $z \sim 0.35$.  BOSS aims to measure spectra (and thus redshifts)
for a sample of 1.5 million galaxies extending to $z = 0.8$ over
10,000~deg$^2$, to use the baryon oscillation feature 
to make a 1\% measurement of the angular diameter distance at 
$z=0.35$ and a separate uncorrelated 1\% measurement at $z=0.6$.  
In addition,
150,000 quasars with $z > 2.15$ will be observed 
to measure the clustering of the Lyman-$\alpha$ forest, 
and thus to determine the
baryon oscillation scale at $z \sim 2.5$, an epoch before dark energy
dominated the expansion of the universe.  

The samples of galaxies and quasars needed to carry out this program
are significantly fainter than those targeted in SDSS-I and SDSS-II
\citep{eisenstein01,strauss02,richards02}, and have a higher density
on the sky.  The SDSS spectrographs and supporting infrastructure were
extensively rebuilt to increase throughput and observing efficiency,
as described in detail in \citet{smee12}.  In particular: 
\begin{itemize} 
\item The optical fibers, which bring light from the focal plane to
  the spectrographs, subtended $3$\arcsec\ on the sky in \mbox{SDSS-I/II}.  Given
  the smaller angular size of the higher redshift BOSS galaxy targets,
  the fibers now subtend $2$\arcsec. 
\item The number of fibers was increased from 640 to 1000. 
\item New high-throughput volume phase holographic (VPH) gratings were installed.
\item The optics have been replaced, with improved throughput. 
\item The CCDs were replaced, with improved response at both the blue and red limits.
\end{itemize}

The resulting spectra are broadly similar to those of \mbox{SDSS-I/II}, but
have significantly higher signal-to-noise ratio (S/N) at a given fiber
magnitude.  While the resolution as a function of wavelength is
similar, the spectral coverage is significantly broader, from 3600\AA\
to 10,400\AA.  Finally, the target selection algorithms for galaxies
\citep{padmanabhan12b} and quasars \citep{ross12} are significantly
different from the equivalent for \mbox{SDSS-I/II}, given the rather
different scientific goals. 

The design of the BOSS survey itself is described in detail in
\citet{dawson12}.  First baryon oscillation results from the DR9
galaxy sample may be found in \citet{anderson12} and references
therein, and the first analysis of the clustering of the Lyman
$\alpha$ forest from BOSS quasar spectra is found in \citet{slosar11}.

\subsection{BOSS Main Survey Targets}

There are four broad categories of targets on the BOSS plates:
galaxies (\S\ref{sec:galaxies}; see \citealt{padmanabhan12b}), quasars
(\S\ref{sec:quasars}; see \citealt{ross12}), ancillary targets
(\S\ref{sec:ancillary}), and standards and
calibrations \citep{dawson12}.  

\subsubsection{Galaxies}
\label{sec:galaxies}

The \mbox{SDSS-I/II} Legacy survey targeted galaxies in two
categories: a magnitude-limited sample of galaxies in the $r$ band
\citep{strauss02}, with a median redshift of $z \sim 0.10$, and a
magnitude- and color-limited sample of fainter galaxies designed to
select the most luminous red galaxies (LRG) at each redshift
\citep{eisenstein01}; the LRG sample is approximately volume-limited
to $z \sim 0.38$, and includes galaxies to $z \sim 0.55$.  BOSS aims
to measure large-scale clustering of galaxies at higher redshifts and
at lower luminosities (to sample the density field at higher space
density), and thus targets significantly fainter galaxies.

The galaxy target selection algorithm is described in detail in
\citet{padmanabhan12b}.  In brief, it uses the DR8 imaging catalog
to select two categories of
objects using colors that track the locus of a passively evolving
galaxy population with redshift \citep{2009MNRAS.394L.107M}.  The ``LOWZ'' subsample,
containing about a quarter of all galaxies in BOSS, 
targets galaxies with $0.15 < z < 0.4$ with colors similar to LRGs, but
with lower luminosity; the space density of LOWZ galaxies is about 2.5
times that of the \mbox{SDSS-I/II} LRG sample.  The constant-mass or ``CMASS'' sample,
containing three times more galaxies than LOWZ, 
is designed to select galaxies with $0.4 < z < 0.8$. The rest-frame color
distribution of this sample is significantly broader than that of the
LRG sample, thus CMASS contains a nearly complete sample of massive
galaxies above the magnitude limit of the survey. 
The LOWZ and CMASS samples together give a very roughly
volume-limited sample, with space density of order 
$3 \times 10^{-4}\,(h/{\rm Mpc}^3)$ to $z \sim 0.6$, 
and a tail to $z \sim 0.8$. 
In practice, it is somewhat difficult to select objects at $z = 0.45$ 
as the 4000\AA\ break falls between the $g$ and $r$ bands.
The space density of the sample at that redshift is consequently 25\% lower. 

The CMASS sample includes a ``SPARSE'' extension in color space, to
better understand incompleteness in the CMASS sample and to sample a
population of fainter, bluer, and less massive galaxies. The galaxies were
selected by extending the CMASS color-magnitude cut, and are sub-sampled at
5 galaxies deg$^{-2}$. 

As described in \citet{padmanabhan12b}, there was an error in the
implementation of the LOWZ sample for the early BOSS data (plate
numbers 3987 and less); these data should be excluded from any
analysis which requires a uniform LOWZ sample.  

The BOSS galaxy sample extends about half a magnitude fainter than the
\mbox{SDSS-I/II} 
LRG sample, and thus the S/N of the spectra tend to be lower, despite
the higher throughput of the spectrographs.  Nevertheless, in DR9 
the vast majority of the galaxy targets are confirmed galaxies with
confidently measured redshifts:
95.4\% of all CMASS targets and 99.2\% of all LOWZ targets.
The 4.6\% of unsuccessful galaxy redshifts for CMASS targets are
mostly erroneously targeted red stars.
As described in \S\ref{sec:galspec}, the signal-to-noise ratio of the spectra is sufficient that
higher-order quantities (stellar masses, velocity dispersions,
emission-line properties, and so on) can be measured for most objects.

\subsubsection{Quasars}
\label{sec:quasars}

The BOSS Quasar Survey uses imaging data from DR8 \citep{DR8} to 
select its main spectroscopic targets.  The aim is to observe 
$z > 2.15$ 
quasars, as for these objects the Lyman-$\alpha$ forest enters into the spectral
coverage of the BOSS spectrographs.  This is a challenging task, given the
fact that the quasar locus in SDSS color space crosses that of F stars
at $z \sim 2.7$ \citep{fan99}.  
\citet{ross12} give full details on the BOSS quasar target selection methods 
that were used.  
In brief, we implemented and tested a range of methods 
over the commissioning period and the first year of BOSS spectroscopy
(Year One, ending in 2010 July). Quasar targets were selected based on
their optical fluxes and colors, and properties in other bands,
including radio and near infrared.  Unlike the \mbox{SDSS-I/II} Legacy quasar
sample \citep{richards02}, the BOSS quasar selection actively selects
{\em against} quasars with redshifts less than $2.15$ (in particular,
most ultraviolet excess sources).

As the main science goal of the BOSS quasar sample is to probe the
foreground hydrogen in the inter-galactic medium (IGM), priority was
placed on maximizing the surface density of $z>2$ quasars
\citep{mcdonald07, mcquinn11}, rather than engineering the most
homogeneous data set possible.  Thus the full target selection is a
complicated heterogeneous combination of several methods, using
ancillary data sets where available \citep{ross12}.

However, to allow statistical studies of 
quasar physical properties, demographics, and clustering, 
we defined a subsample \citep[called ``CORE'' in
][]{ross12} that will be uniformly 
selected throughout BOSS.  It uses a single selection algorithm, the
extreme deconvolution method (hereafter XDQSO) of \citet{bovy11},
using single-epoch SDSS photometry.  However, we settled on XDQSO only
at the end of Year One, and thus the CORE sample in the first year of
data is {\em not} homogeneous.  CORE targets were allocated at
20~deg$^{-2}$, of which $\sim10$--$15$~deg$^{-2}$ are confirmed
spectroscopically to be quasars at $z>2$.  An additional 20 targets
deg$^{-2}$  
(the ``BONUS'' sample) 
were
selected using a heterogeneous set of selection criteria to maximize
the surface density of high-$z$ quasars; 
of these, $\sim5$ deg$^{-2}$ are found to be quasars at $z>2$.  
In Year One, especially in
the commissioning period, we increased the number density of targets
as we fine-tuned the selection algorithms.  

Finally, given the improved throughput of the BOSS spectrographs and 
extended blue coverage, we
re-observed all previously known $z>2.15$ quasars (most of which were
discovered by \mbox{SDSS-I/II}; see \citealt{DR7Q}) to obtain higher
S/N in the Lyman-$\alpha$ forest. 

Approximately half of the quasar targets
observed in DR9 were confirmed to be quasars, with the remainder
consisting largely of F stars.

All quasar targets, plus all objects spectroscopically identified as
quasars via our automated pipeline, have been visually inspected, and
both automated pipeline results and these visual redshifts and 
classifications are provided in DR9.
The resulting quasar catalog, 
together with measurements of broad absorption lines and damped
Lyman-$\alpha$ systems, will be made public in \citet{paris12}.  
A subsample of BOSS quasar spectra suitable for 
Lyman-$\alpha$ forest analysis ($z \geq 2.15$) will be described in 
\citet{lee12}, which will provide additional products 
such as quasar continua, improved noise estimates, and pixel masks. 

\subsubsection{BOSS Ancillary Targets}
\label{sec:ancillary}

In addition to the main galaxy and quasar programs, roughly 3.5\% of the
BOSS fibers in DR9 were devoted to a series of 25 small ancillary projects, 
each consisting of a few hundred to a few thousand targets.
These programs, described in detail in Appendix A of
\citet{dawson12}, were selected via internal collaboration review, and
cover scientific goals ranging from studies of nearby stars to $z>4$ 
quasars.  The ancillary programs allow fibers to be used that would
otherwise go unplugged in 
regions where the principal targets are more sparse than
average. These spectra are processed with the same pipeline
\citep{schlegel12,bolton12} as all the 
other spectra. 

A particular focus of many of these ancillary programs is 
the roughly 220~deg$^2$ in the Southern Galactic Cap
covered by ``Stripe 82''
($-1.25^\circ < \delta < +1.25^\circ$, $320^\circ < \alpha < 45^\circ$)
that was imaged repeatedly in SDSS \citep{DR6}.  Using
stacked photometry and variability information, for example, the
quasar sample on Stripe 82 is particularly complete 
(e.g., \citealt{palanque11}). 

\subsection{Differences between \mbox{SDSS-I/II} Spectra and \mbox{SDSS-III} BOSS Spectra}

\label{sec:sdss_vs_boss_spectra}

Readers who are familiar with the \mbox{SDSS-I/II} spectra will be able to
use the BOSS spectra quickly, since the twin BOSS spectrographs are
upgraded versions of the original \mbox{SDSS-I/II} spectrographs, as
described above. In addition, the pipelines used to process 
the BOSS spectra 
\citep{schlegel12,bolton12} 
are improved versions of those used in \mbox{SDSS-I/II}.
In this section, we briefly outline the
main differences between the BOSS spectra and the \mbox{SDSS-I/II} spectra.
For more detailed information on the BOSS spectrographs, the reader is
referred to \citet{smee12}, while the BOSS operations are
described in \citet{dawson12}.

The BOSS spectrographs include 1000 fibers in each plate, in
comparison with 640 fibers per plate in \mbox{SDSS-I/II}.  In addition, the
spectral coverage has been increased from $3800$--$9200$~\AA\ to
$3600$--$10,400$~\AA, with the dichroic split between the blue and red
sides occurring at roughly $6000$~\AA\ (as it was in \mbox{SDSS-I/II}).  The
expanded blue coverage means that the \ion{Cd}{1} arc line at
$3610.51$~\AA\ is now included in the wavelength calibration, enabling
a more accurate wavelength solution on the blue end (see the
discussion in \citealt{DR6}).  The median resolution of the BOSS
spectra remains $R = \lambda / \Delta\lambda \approx 2000$ as in
\mbox{SDSS-I/II}, with a similar wavelength dependence 
\citep{smee12}; 
the resolution ranges from $R \approx 
1500$ at $3700$~\AA, to $R \approx 2500$ at $9500$~\AA.

In addition, the diameter of the spectroscopic fibers in BOSS has been decreased in
size from $3\arcsec$ to $2\arcsec$.  While this improves the
S/N for point-like objects and the smaller galaxies
targeted by BOSS due to decreased sky background relative to the source signal, 
the smaller fiber size affects the spectrophotometry
for galaxies, and is more subject to differential chromatic aberration
and seeing effects.  As in \mbox{SDSS-I/II}, the spectrophotometry is tied to
the PSF photometry of stars on each plate.  In \mbox{SDSS-I/II}, the RMS 
scatter between the PSF photometry and synthesized photometry from the
calibrated spectra was of order 4\% \citep{DR6}; with BOSS, it is
closer to 6\% (\citealt{dawson12}, but see the discussion below about quasar
spectrophotometry).
The photometric catalog released in DR8 and DR9
(\S\ref{sec:distribution}) provides the 2\arcsec\ photometry (termed
\mbox{FIBER2MAG}) for each object to complement 3\arcsec\ photometry
(FIBERMAG).  

\begin{figure}
\plotone{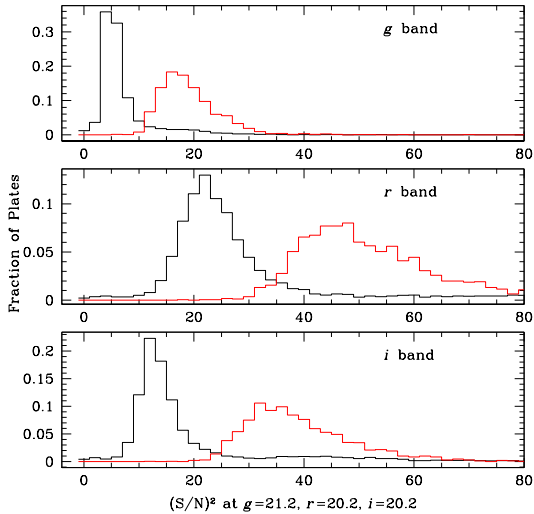}
\caption{S/N per pixel distribution of DR9 BOSS plates (red),
  compared with the equivalent for DR7 \mbox{SDSS-I/II} plates (black).  The
  quantity shown is the square of the S/N, measured
  at a fiducial fiber magnitude.  In SDSS-I and SDSS-II, these
  fiducial magnitudes differ somewhat (and the flux is measured
  through a $3$\arcsec\ fiber, not a $2$\arcsec\ fiber); these effects have been
  accounted for in this figure to make a fair comparison.}
\label{fig:boss_snr}
\end{figure}

The more 
sensitive CCDs, improved throughput of the VPH gratings, and improved
optics have further improved the S/N in the BOSS
spectra, enabling the targeting of fainter objects.  For each plate,
the median log S/N per pixel within wavelength regions
corresponding to the SDSS imaging bands $g, r$ and $i$
\citep{fukugita96} is tabulated against the corresponding $2$\arcsec\ fiber
magnitude.  A line of slope 0.3 is fit to this line, and the intercept
at the fiducial magnitudes of $g=21.2, r=20.2$ and $i=20.2$ is
noted.
This quantity is compared for SDSS DR7 and BOSS plates in 
Figure~\ref{fig:boss_snr}.  
The median exposure times of BOSS DR9 plates (1.5 hours) are only 70\%
longer than those in \mbox{SDSS-I/II} (0.89 hours), 
but due to the instrument upgrades the resulting (S/N)$^2$ values of the BOSS
spectra are more than twice those in \mbox{SDSS-I/II} at the same magnitude. 

Because one of the stated goals of the BOSS survey is to study the
Lyman-$\alpha$ forest absorption in quasars, efforts have been made to
improve the S/N at the blue end of the BOSS objects
targeted as quasars.  In particular, the focal plane of the SDSS
telescope was designed to be in focus for BOSS at $\sim 5400$~\AA, whereas the
$z \sim 2.5$ Lyman-$\alpha$ forest lies at $\lambda \lesssim
4000$~\AA, a wavelength that will be out of focus and offset radially due to
differential chromatic aberration.  To correct for this, we have offset the
quasar target fibers in both the radial and axial directions to
maximize the throughput at $\lambda \sim 4000$~\AA.  The radial offset was
implemented by drilling the quasar plug holes at slightly different
positions (depending on the assumed hour angle at which the plate will
be observed), whilst in the axial direction we have introduced thin washers to the
plug holes on the fiber side of the plates, with thicknesses of 175
and 300~micron in the regions spanning 1.02--1.34~deg and
1.34--1.50~deg radially from the plate center, respectively
\citep{dawson12}.  These offsets are tabulated in the \code{ZOFFSET} and
\code{LAMBDA\_EFF} flags in the survey data (\S\ref{sec:distribution}). 

The current pipeline flux calibration \citep{schlegel12} does not take
these fiber offsets into account, therefore the spectrophotometry of
the objects in the quasar targets is 
biased toward bluer colors, with excess flux relative to the SDSS
imaging data at $\lambda<  4000$ \AA\ and a flux decrement at longer
wavelengths \citep{paris12}.  
We have measured the mean difference between spectrophotometric and
imaging magnitudes for those objects targeted as quasars but that turned
out to be stars\footnote{We exclude quasars from this comparison to
  avoid introducing intrinsic quasar variability between the time the
  photometry and spectroscopy were carried out into the comparison
  between the two different magnitudes.}
--  
the values are ($0.11\pm0.24, 0.16\pm0.29,
0.24\pm0.33$)~mag in ($g$, $r$, $i$).  Objects observed at higher
airmass show larger offsets.

Quasars targeted solely as part of ancillary programs were not subject to
these offsets, and thus their spectrophotometry should show no
significant bias.  
Of course, these objects will have reduced S/N in the blue.  
However, some quasars targeted in ancillary programs were also targets
in the main CORE or BONUS samples; these ancillary quasars 
do have the washer offsets applied (at least after MJD 55441, when the
washers started to be applied; see \S\ref{sec:boss_changes} below).  

DR9 includes new BOSS observations of objects observed with the
previous spectrograph in \mbox{SDSS-I/II}.  This includes 4,074 galaxies;
16,967 quasars (mostly specifically re-targeted to obtain better
Ly~$\alpha$ forest measurements); and 7,875 stars.  The repeated galaxy and
star observations confirm that the redshift scales are consistent
within a few \kms. However, due to an updated
set of quasar templates in the BOSS pipeline \citep{bolton12}, quasar
redshifts are 175 \kms\ higher in the median in BOSS than in
\mbox{SDSS-I/II}.  
The limitations of the quasar redshifts in previous data releases
were highlighted by \citet{hewett10} in a reanalysis of DR6
quasar redshifts.
While the new templates are designed to more fully represent the
range of quasars found, obtaining accurate redshifts remains
challenging because of the uncertainty in the relative velocity
offsets of different emission lines from the rest frame of the quasar
host galaxy system. 
See
\citet{paris12} and \citet{bolton12} for a discussion of the details
and caveats of quasar redshift determination in DR9. 

\begin{figure*}
\plotone{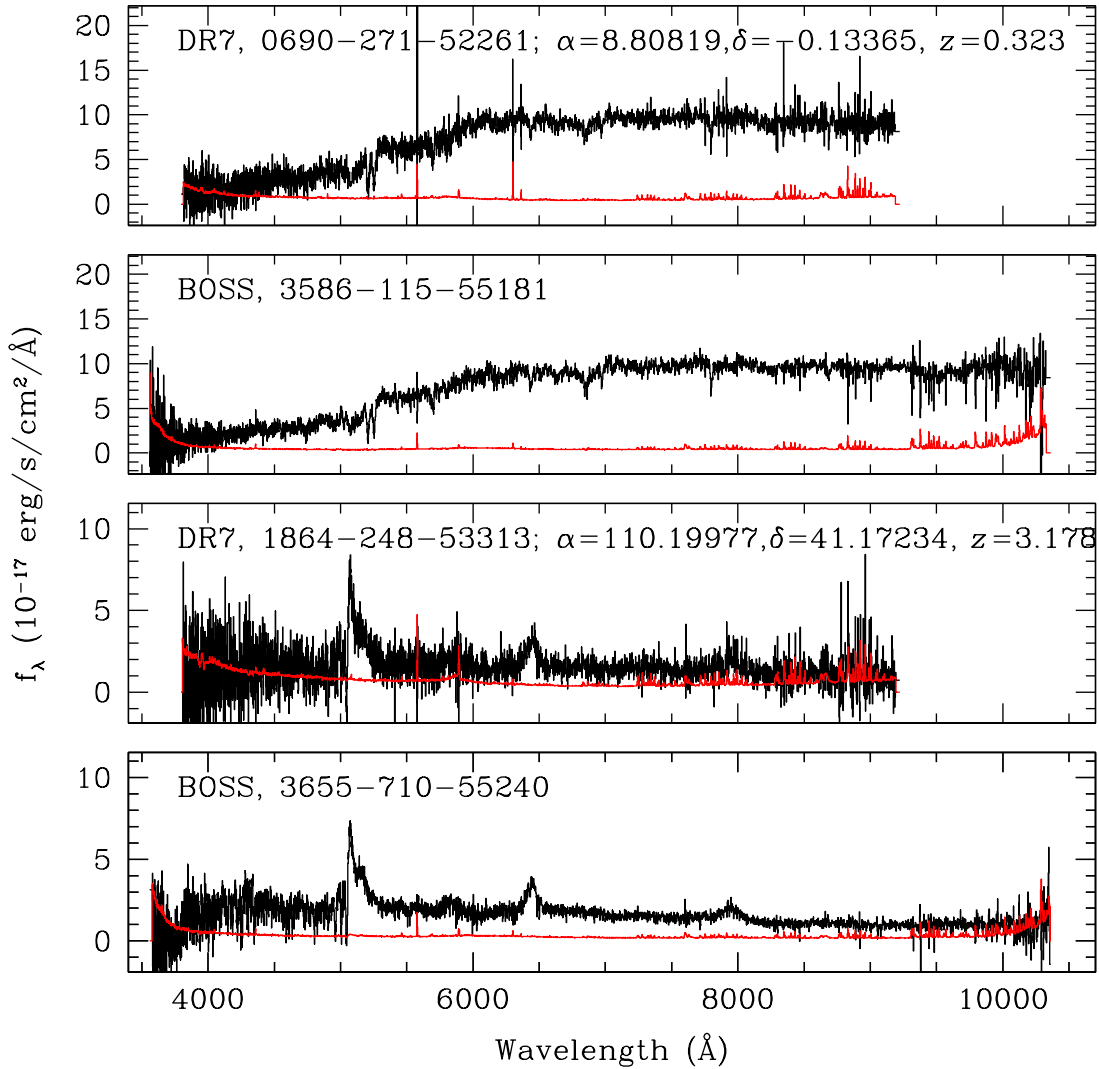}
\caption{A galaxy (upper panels) and a quasar (lower panels) that were observed in
  both \mbox{SDSS-I/II} (as released in DR7) and BOSS.  These spectra
  are unsmoothed.  In addition to
  the extended BOSS wavelength coverage from 3600 to 10,400~\AA, the
  estimated noise per pixel (red line) is lower at every wavelength for the BOSS spectra, particularly at the
  red and blue ends of the spectrum.  This is consistent with the
  higher S/N of the BOSS spectra shown in the distributions in
  Fig.~\ref{fig:boss_snr}.  Because the SDSS-I/II spectra are observed
  through $3''$ fibers, while the BOSS spectra use $2''$ fibers, one
  does not expect the galaxy spectra to be identical.  
}
\label{fig:boss_sdss_gal_qso}
\end{figure*}

Figure~\ref{fig:boss_sdss_gal_qso} shows spectra of a galaxy and
a quasar, observed both with \mbox{SDSS-I/II} and BOSS. This figure illustrates
the greater wavelength coverage
and the significantly higher S/N of the BOSS spectra for observations of the same object. 

\subsection{Quantities Derived from Galaxy Spectra}
\label{sec:galspec}

The spectroscopic pipeline \citep{bolton12} initially classifies all
spectra without regard to its imaging data.  That is, each object is
tested against galaxy, quasar, and stellar templates, regardless of
how it was targeted.  However, in BOSS, we found that galaxy targets were
often incorrectly matched to quasar templates with unphysical fit
parameters, e.g., negative 
coefficients causing a quasar template emission feature to fit a galaxy
absorption feature.  Thus, for galaxy targets in BOSS, the best classification
and redshift are selected only from the fits to the galaxy and star
templates.  The resulting quantities are listed with the suffix
\code{\_NOQSO} in the DR9 database.  Results without this template
restriction are also made available.

In addition, we have computed a variety of derived quantities from the
galaxy spectra following the spectroscopic pipeline, applying stellar
population models to derive stellar masses, emission-line fluxes and
equivalent widths, stellar and gas kinematics and velocity
dispersions \citep{chen12,maraston12,thomas12}.

Each of the stellar population models is
applied to all objects that the spectroscopic pipeline calls a galaxy
with a reliable and positive definite redshift (i.e.,
\code{CLASS\_NOQSO}=``GALAXY'' and \code{ZWARNING\_NOQSO}=0 and
\code{Z\_NOQSO} $>$ \code{Z\_ERR\_NOQSO} $>$ 0; see
\citealt{bolton12}).  

\begin{itemize}
\item Portsmouth spectro-photometric stellar masses~\citep{maraston12} are
  calculated using the BOSS spectroscopic redshift, \code{Z\_NOQSO}, and
  $u,g,r,i,z$ photometry by means of broad-band spectral energy distribution
  (SED) fitting of population models. Separate calculations are carried out with
  a passive template and a star-forming template, and in each case for both
  \citet{salpeter55} and \citet{kroupa01} initial mass functions, 
  and for stellar evolution with and without stellar mass loss. 
  Templates are based on \citet{2005MNRAS.362..799M} and 
  \citet{2009MNRAS.394L.107M} for the star-forming and passive stellar 
  population models, respectively.  
  In order not to underestimate stellar mass, 
  internal galaxy reddening is not included in the Portsmouth SED fitting procedures used in DR9. 
  Reddening for individual galaxies may, however, be computed via the 
  Portsmouth emission-line flux calculations (see below).
\item Portsmouth emission-line fluxes and equivalent widths, and stellar
  and gas kinematics \citep{thomas12}, 
  are based on the stellar population synthesis
  models of \citet{2011MNRAS.418.2785M}
  applied to BOSS spectra using an
  adaptation of the publicly available Gas AND Absorption Line
  Fitting~\citep[GANDALF;][]{2006MNRAS.366.1151S} and penalized PiXel
  Fitting~\citep[pPXF;][]{2004PASP..116..138C}. 
\item Wisconsin stellar masses and velocity dispersions are derived from the
  optical rest-frame spectral region (3700--5500\AA) using a 
  principal component analysis (PCA) method \citep{chen12}.
  The estimation is based 
  on a library of model spectra generated using the single stellar population 
  models of \citet{2003MNRAS.344.1000B} assuming a \citet{kroupa01} initial mass function, and with a broad range of star-formation histories,
  metallicities, dust extinctions, and stellar velocity dispersions.
\end{itemize}

The different stellar mass estimates for BOSS galaxies encompass calculations based
on different stellar population models (\citealt{2003MNRAS.344.1000B}
for Wisconsin, and \citealt{2005MNRAS.362..799M} for Portsmouth),
different assumptions regarding galaxy star formation histories, and
multiple choices for the initial mass function and stellar mass-loss
rates, and each method focuses on a different aspect of the available
imaging and spectroscopic data. 
The Portsmouth SED fitting focuses on broad-band colors and BOSS redshifts, the Wisconsin PCA analysis on considering the full spectrum, while the Portsmouth emission-line fitting focuses on specific regions of the spectrum that contain specific information on gas and stellar kinematics.
The uncertainty in the Wisconsin spectral PCA results generally decreases with increasing spectrum S/N, whereas the Portsmouth SED-fit results provide a wider choice of stellar population models relevant to BOSS galaxies.  
The array of choices allows consistent comparisons with the literature and future surveys. 
A detailed comparison between the Portsmouth SED and the Wisconsin spectral PCA calculations is discussed in \citet[][Appendix A]{maraston12}.

The Galspec
product~\citep{2003MNRAS.341...33K,2004MNRAS.351.1151B,2004ApJ...613..898T}
provided by the Max Planck Institute for Astrophysics and the Johns
Hopkins University (MPA-JHU) introduced in DR8 is maintained for SDSS-I/II
galaxies, but is not available for \mbox{SDSS-III} BOSS spectra.
The Portsmouth and Wisconsin stellar population model algorithms are new
to DR9 and currently available only for \mbox{SDSS-III} BOSS spectra. 
However, \citet{chen12} and \citet{thomas12} each found that a comparison of their respective techniques (Wisconsin PCA, and Portsmouth emission-line) to the SDSS-I/II MPA-JHU demonstrated consistent results with the values for a set of SDSS galaxies from DR7.

\subsection{Changes in BOSS Spectrographs and Survey Strategy}
\label{sec:boss_changes}

\begin{figure}
\plotone{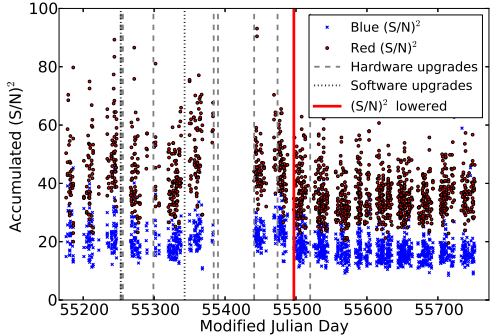}
\caption{
Accumulated signal-to-noise ratio squared per pixel at a fiducial
magnitude on each plate, plotted as a function of time for the BOSS
survey data presented in DR9 for all completed plates marked as good. 
The blue (S/N)$^2$ is the average of the signal in blue cameras of the spectrograph for an object with $g=21.2$, while the red (S/N)$^2$ is the average of the red cameras of the spectrograph for an object with $r=20.2$.
Survey-quality data began at MJD 55171. Changes in survey strategy,
hardware, and guider software (Table~\ref{table:boss_changes}) are 
indicated with vertical lines.
The mean signal-to-noise ratio per plate dropped significantly after
the requirements for exposure depths were reduced on MJD 55497
(\S\ref{sec:boss_changes}).  
  The large gap is the 2010 summer shutdown.  The smaller gaps are the times of bright moonlight when BOSS does not observe.
}
\label{fig:boss_changes}
\end{figure}

\begin{deluxetable*}{lll}
\tablecaption{BOSS Survey Changes\label{table:boss_changes}}
\startdata
Date & MJD\tablenotemark{a} & Change \\ \hline
2009 Aug 28 &55071 & Earliest BOSS commissioning data available in SAS \\
2009 Dec 06 &55171 & Beginning of survey-quality data \\
2010 Feb 26 &55253 & Installed mask in the central optics to eliminate a secondary light path that was directly imaged onto the CCD \\
2010 Feb 26 &55253 & Guider improvements \\
2010 Feb 26 &55253 & CCD dewars adjusted for better focus \\
2010 Mar 01 &55256 & Installed a collimator mask to remove light being reflected off of the slithead and re-imaged onto the CCD. \\
2010 Apr 12 &55299 & R2 CCD replaced \\
2010 May 28 &55343 & Field acquisition and calibration efficiency improvements \\
2010 Jul  7 &55384 & CCD positions adjusted inside dewars for better focus \\
2010 Jul 13 &55390 & R1,R2 CCD change from 1- to 2-phase slightly changed effective pixel size \\
2010 Sep 02 &55441 & Washers for quasar targets, some plates \\
2010 Oct 05 &55474 & Washers for quasar targets, all plates \\
2010 Oct 28 &55497 & Changed (S/N)$^2$ thresholds and target selection \\
2010 Nov 20 &55520 & B1 triplet lenses replaced 
\enddata
\tablenotetext{a}{All data taken on and after the given MJD include the respective change.}
\end{deluxetable*}

While commissioning of the BOSS spectrographs was completed in early
December 2009, we continued to make a series of improvements and
changes to the spectrographs, the observing system, and the exposure
depths.  In this section, we outline those changes that affect the DR9
data.  The effects on the quality of the resulting spectra due to these
changes are subtle, but the reader interested in detailed comparisons
of the BOSS data as a function of time should be aware of them.  

BOSS observes spectra with 15 minute exposures which are repeated
until the summed signal-to-noise squared per pixel, (S/N)$^2$, reaches a given
threshold in each of the four spectrograph cameras
(B1, B2, R1, R2 for the blue and red arms of spectrographs 1 and 2).
A quick-look pipeline runs after each exposure to estimate the
accumulated (S/N)$^2$ in near real-time and a plate is exposed again
until given (S/N)$^2$ thresholds are reached.

For the first year of the survey BOSS conservatively 
observed a little deeper than believed necessary and planned
on re-evaluating and updating these (S/N)$^2$ thresholds for future years.
After the first year of observations, it became clear that that we were not
covering the sky sufficiently quickly to reach our goal of 10,000 deg$^2$ 
by the end of the survey in Summer 2014.  
BOSS thus conducted a review of the fiducial (S/N)$^2$ thresholds
needed to optimize both survey speed and spectroscopic completeness.  
The decision was made to lower the (S/N)$^2$ thresholds 
and impose a more restrictive cut
on the galaxy surface brightness faint limit.
On MJD 55497 the (S/N)$^2$ thresholds were reduced
from $>16$ to $>10$ for the blue spectrograph cameras (for $g=22$) and
from $>26$ to $>22$ for the red spectrograph cameras (for
$i=21$).\footnote{These values of (S/N)$^2$ are as measured by the
  quick reductions done of each exposure immediately after it is
  taken.  The full reductions have a moderately higher (S/N)$^2$.  
  The full pipeline also uses a different set of fiducial magnitudes 
  for tracking (S/N)$^2$: $g=21.2$~mag, $r=20.2$~mag, and $i=20.2$~mag.  
  It is these full pipeline (S/N)$^2$ values that are shown in Figure~\ref{fig:boss_snr}.}
At the same time, the CMASS target selection limiting magnitude
was changed from \code{IFIBER2MAG} $<21.7$ to $<21.5$.  There is a
very slight change in spectroscopic survey completeness after this
date.  Further details are provided in
\citet{dawson12}, \S{5}.

Improvements to the guider software were made on MJD 55253,
leading to better guiding and thus improved throughput.  
Improvements to the field acquisition software and the efficiency of calibration observations were made on MJD 55343 and resulted in reduced observing overheads and a larger fraction of open-shutter time.

Table~\ref{table:boss_changes} summarizes a series of hardware changes
that further improved throughput and image quality and reduced
scattered light.  This allowed us to reach the fiducial (S/N)$^2$ in
the spectra in fewer exposures.  
Air bubbles had developed in the oil interfaces between the B1 triplet
lenses, reducing throughput and causing scattered light.  These  were
replaced on MJD 55520. The triplet lenses for the other spectrograph
arms have also been replaced, but after the 2011 July date that
marks the end of DR9. 
The R2 CCD was replaced on MJD 55298 due to a hardware failure.
The R1 and R2 CCD clocking was changed from 1- to 2-phase for charge collection on MJD 55390.
The use of washers to optimize (S/N)$^2$ for quasar targets began on
MJD 55441 and was fully implemented for all CORE and BONUS quasar
targets starting MJD 55474. Finally, we did two rounds of adjusting
the focus of the CCDs in their dewars, further improving the
throughput. 

\section{Fixed and Improved Astrometry}
\label{sec:astrometry}

\begin{figure*}
\plottwo{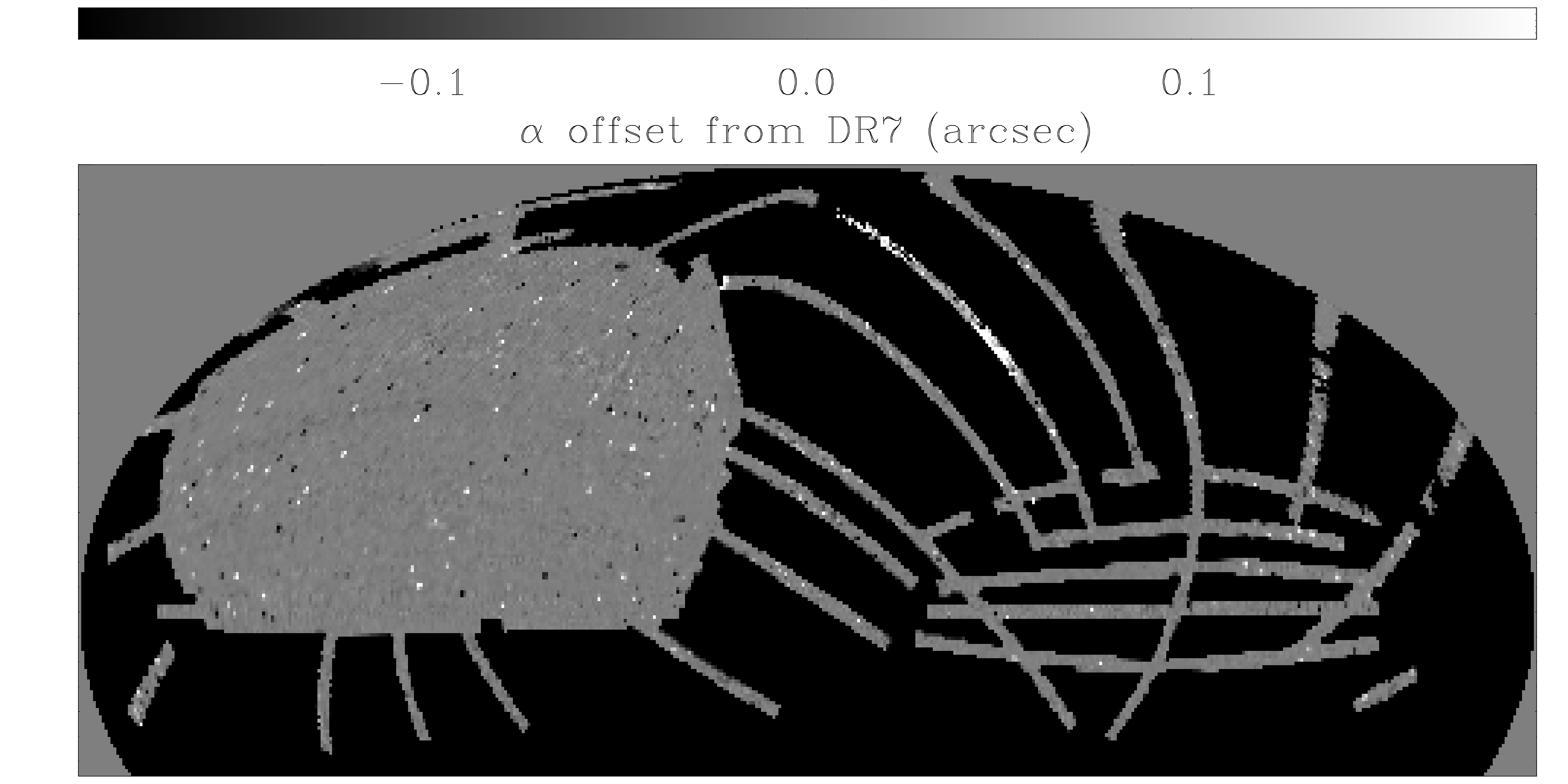}{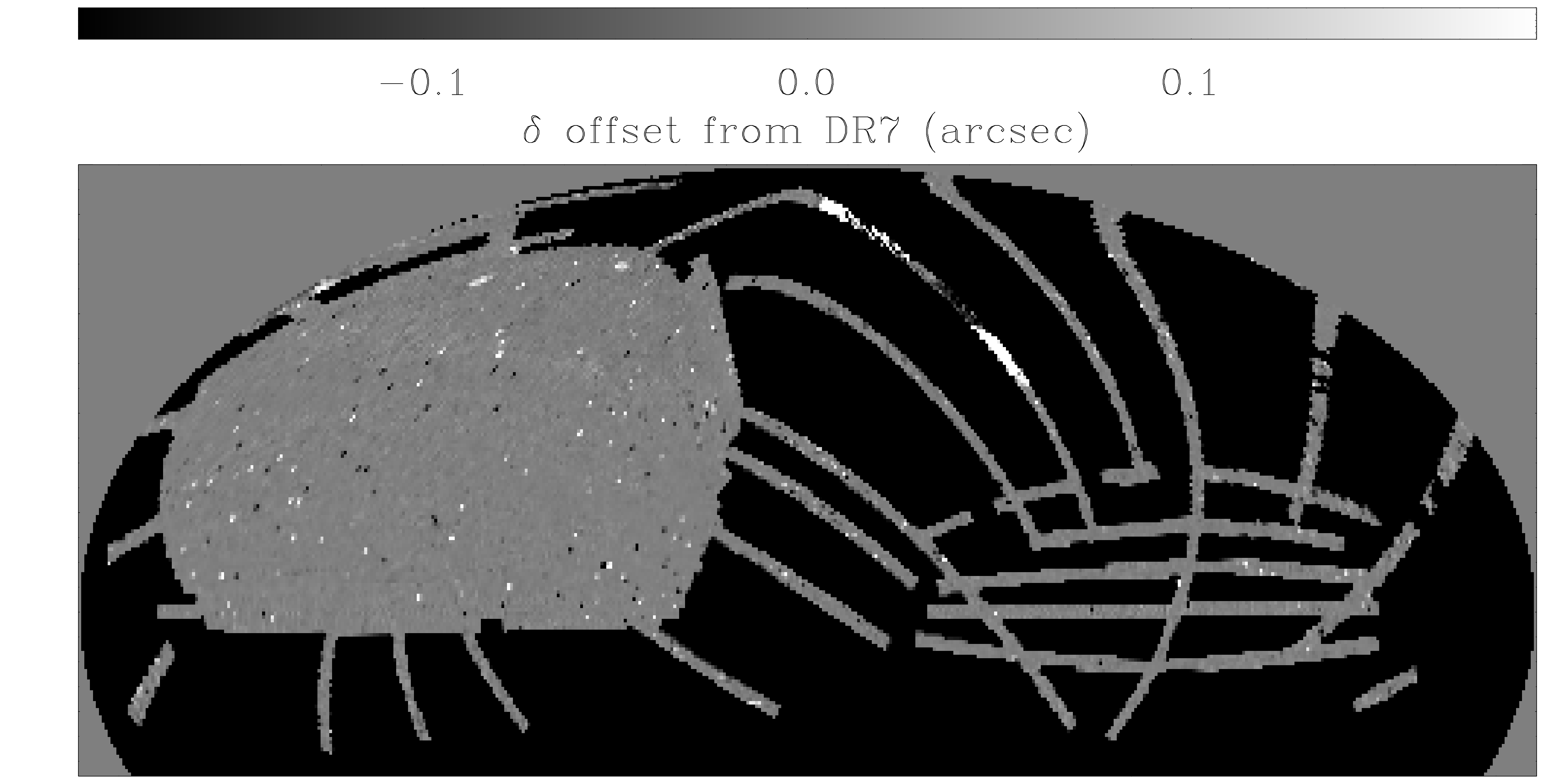} \\
\plottwo{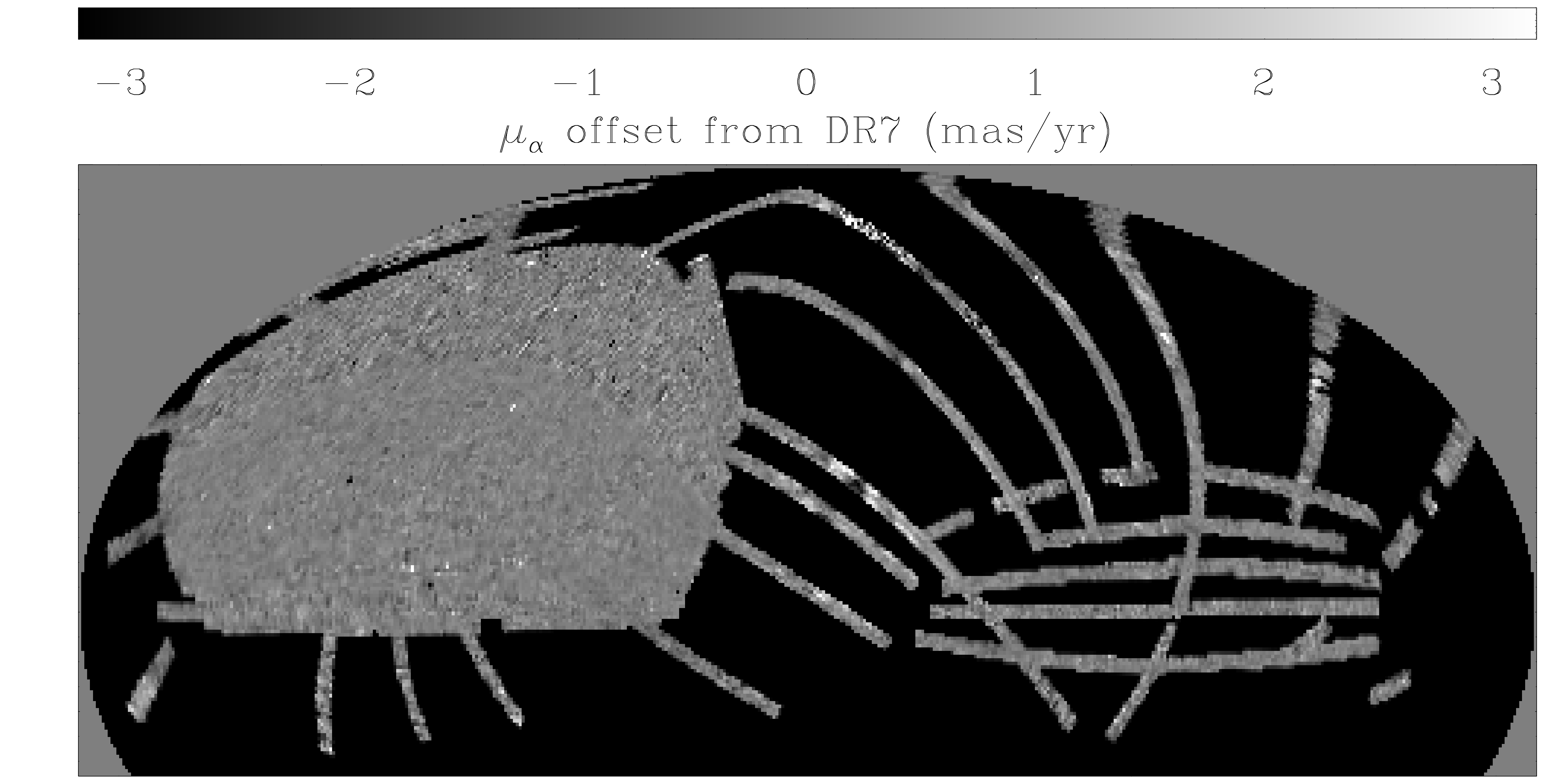}{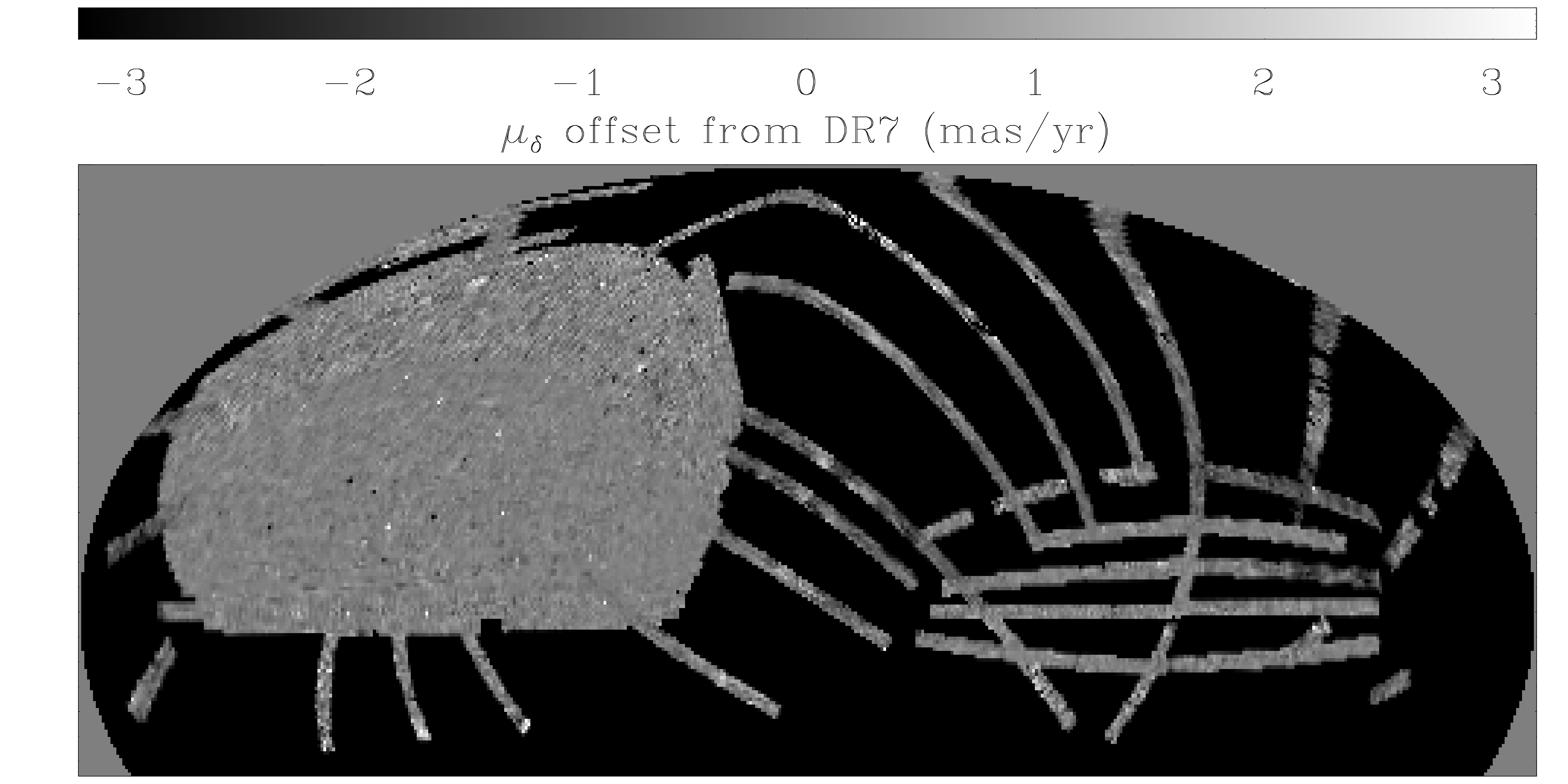}
\caption{ Astrometric and proper motion comparison of DR9 to DR7,
  plotted in equatorial coordinates. The top row shows the difference
  in right ascension (left) and declination (right) of objects
  matched between the two data 
  releases, and the bottom row shows the differences in their
  proper motions.  In the top row, 
  the DR7 and DR9 astrometry agree over most of the area, with the
  exception of a 
  handful of spots, all due to errors in the DR7 astrometry.  In the
  bottom row, DR7 and DR9 proper motions agree
  over virtually all of the high Galactic latitude areas. 
  At low Galactic latitudes there are substantial shifts, caused by
  errors in DR7 due to mistakes in star/galaxy separation affecting
  the proper motion estimates.}
\label{fig:dr9_vs_dr7_astrometry}
\end{figure*}

\begin{figure*}
\plottwo{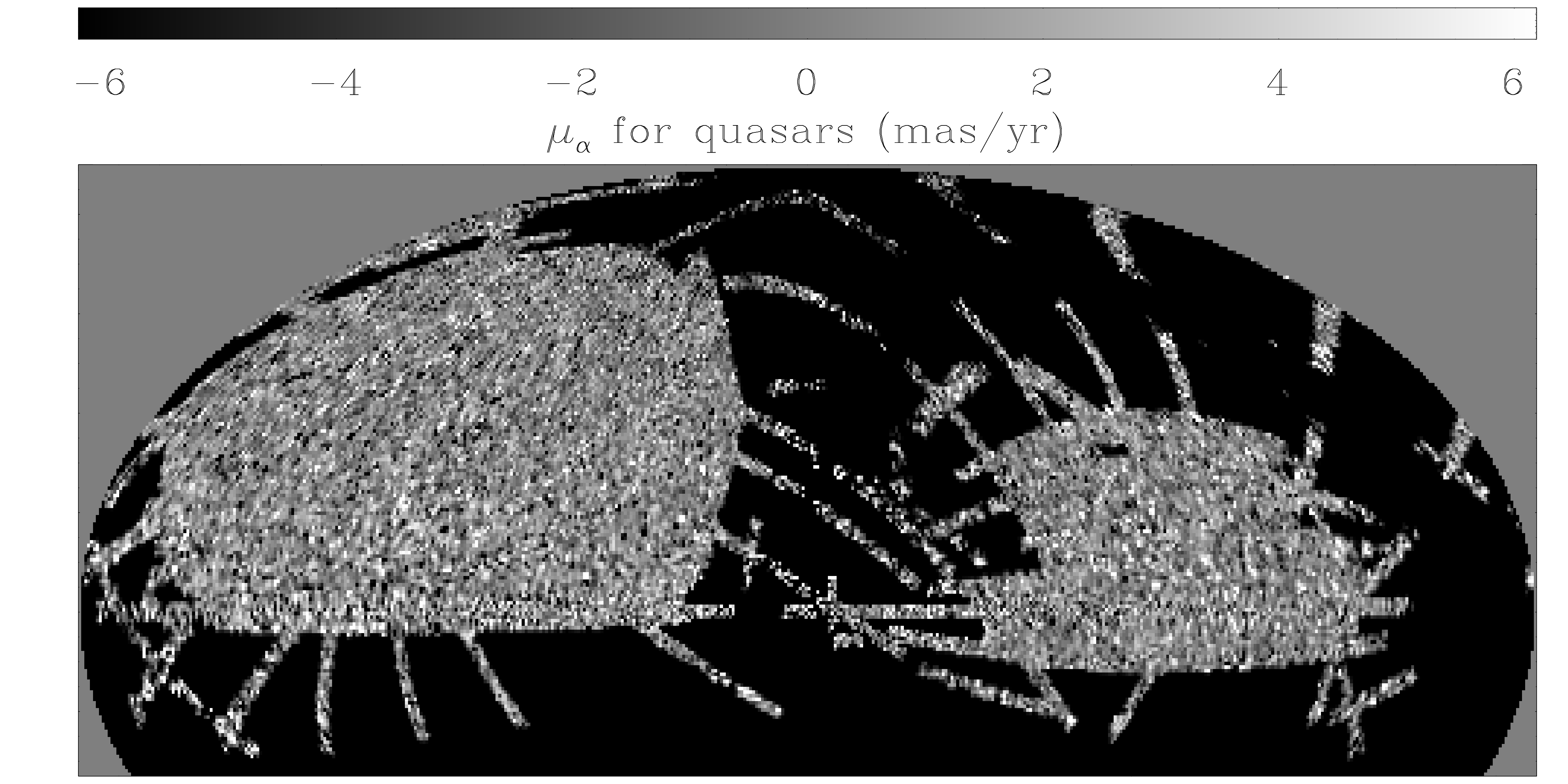}{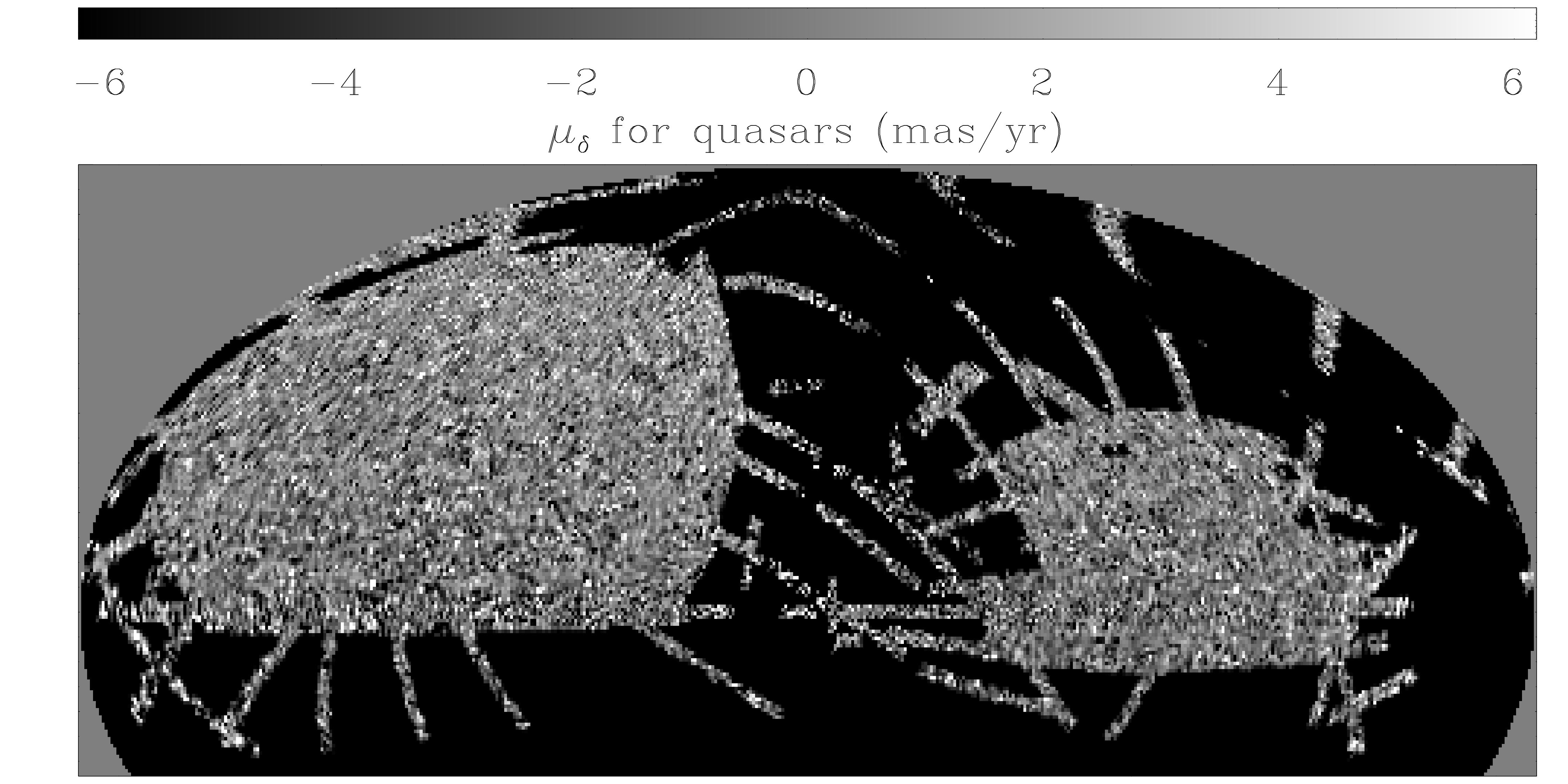}
\caption{ The DR9 proper motions of photometrically-selected $z<2$
  quasars \citep[as classified by][]{bovy11}.  These motions are
  nearly consistent with zero, with a slight offset in $\mu_\delta$ at
  low declination, possibly due to errors in differential refraction
  corrections in USNO-B for these very blue objects (see
  \citealt{bond10a}).  }
\label{fig:dr9_astrometry}
\end{figure*}

The DR8 imaging suffered from several errors in the astrometric
calibration, as described in an erratum published shortly after the
DR8 release~\citep{DR8_erratum}.\footnote{These errors do not appear
  in the DR7 and earlier releases.}  
These errors have been corrected in DR9, and the resulting astrometry
and proper motions are improved relative to both DR7 and DR8.\footnote{While the FITS
  images distributed as part of the Science Archive
  Server, \url{http://data.sdss3.org/sas/dr9/}, are identical to DR8
  on a pixel-by-pixel basis, the FITS image metadata (in particular,
  the World Coordinate System headers) have been changed to match the
  revised astrometry in DR9.}

The issues with the DR8 astrometry were, in brief:
\begin{itemize}
\item Northward of $+41^\circ$ declination there was an offset of
  250~mas introduced by switching from the Second US Naval Observatory CCD
  Astrograph Catalog (UCAC2; \citealt{zacharias04}) to the
  United States Naval Observatory (USNO)-B catalog \citep{monet03} at this declination. 
\item Color terms were not used in calculating CCD position to sky
  position, introducing systematic errors of 10--20~mas.
\item UCAC2 proper motions were not applied correctly, introducing
  further errors of order 5--10~mas.
\item Stellar positions were always measured in the $r$-band photometry, even
  if the $r$-band was saturated or had a lower S/N detection than other filters.
  For faint objects this increases the statistical uncertainty for the
  measurement, but for $r$-band saturated objects the difference can be
  as much as 100~mas between using $r$-band positions and those in non-saturated
  filters. 
\end{itemize}

All of these issues have been corrected for DR9.  The discovery of the
mistakes in DR8 prompted the development of a new set of astrometric
quality-assurance metrics that are fully described in the \mbox{SDSS-III} DR9
datamodel.\footnote{\url{http://data.sdss3.org/datamodel}}

With these problems corrected, the DR9 astrometry fixes errors in both
DR8 and DR7. In particular, DR7 contained very large errors in a
handful of runs (3358, 4829, 5960, 6074, and 6162) that are corrected
in DR9 (the most prominent of these is the black and white arc in the
upper central region of the top panels of
Figure~\ref{fig:dr9_vs_dr7_astrometry}).

Proper motions in DR9 are similarly improved relative to DR8 and DR7.
As Figure \ref{fig:dr9_vs_dr7_astrometry} shows, they are mostly
unchanged in the mean at high Galactic latitudes. However, the
corrected color terms in the astrometry have fixed a small fraction of
objects with outlying proper motions in the DR9 relative to DR8 (this
error did not affect DR7 or earlier). 
Furthermore, at low Galactic latitudes DR7 had
some large offsets caused by star-galaxy separation errors.  Proper
motions are measured with respect to a reference frame of stationary
galaxies, so stellar contamination in the galaxy sample can
systematically affect the proper motion estimates. In DR7, errors
in star-galaxy separation (in particular in photometric rerun 648)
caused the galaxy sample to have significant stellar contamination,
leading to systematic errors in the proper motions.  DR9 fixes this problem.

The proper motions can be independently tested by looking at the
proper motions of photometrically identified low redshift quasars,
which are easy to select and should have zero proper motions.  Figure
\ref{fig:dr9_astrometry} shows the proper motions of the low redshift
quasars as selected by \citet{bovy11}. These show very little
systematic offset from zero, except for a small shift in $\mu_\delta$
at low Declination.  This offset is further described by
\citet{bond10a} in the context of the DR7 proper motions, and could be
due to small differential refraction correction issues in USNO-B for
these very blue objects (and is therefore likely not relevant to the
proper motions of typical stars).

\section{Improvements in the SEGUE Stellar Parameter Pipeline for DR9}
\label{sec:SSPP}

The SEGUE Stellar Parameter Pipeline
\citep[SSPP;][]{lee08a,lee08b,allendeprieto08,smolinski11} utilizes
multiple approaches to estimate effective temperature ($T_{\rm
  eff}$), surface gravity (log $g$), and metallicity ([Fe/H]) from
stellar spectra. Each method is optimized for a certain range of
stellar color $(g-r)$
and S/N, and is measured over a range of wavelengths determined to
deliver the best estimate of each parameter. 
The SSPP is designed to obtain reliable results for stars targeted
as part of the SDSS-II SEGUE and \mbox{SDSS-III} SEGUE-2 surveys \citep{rockosi12}.
With each SDSS data release the SSPP has been refined and modified 
to provide more accurate estimates of the
stellar atmospheric parameters.
Here
we briefly highlight major changes and improvements made since the DR8
public release that are used for the DR9 data.  

A sample of 126 high-resolution spectra of SDSS/SEGUE stars, taken
with Keck, Subaru, the Hobby-Eberly Telescope and the Very Large
Telescope, have 
been analyzed in a homogeneous fashion, and a new set of stellar
parameters were obtained from this analysis
\citep{allendeprieto08, smolinski11}.
The sample covers $4000<T_{\rm eff}<7000$~K, $0.0<\log{g}<5.0$, and $-4.0<$~[Fe/H]~$<+0.5$. 
However, this data set contains no metal-poor ([Fe/H]~$<-2.5$) dwarfs or
metal-rich ([Fe/H]~$>0.0$) giants.  
Additional information on this high-resolution sample can be found in 
\citet{allendeprieto08} and \citet{smolinski11}.

The individual methods in the SSPP, in particular estimates of surface gravity
and metallicity, have been thoroughly re-calibrated
based on these new data. The SSPP also adopts a much-improved color 
($g-r$)-temperature relation, the InfraRed Flux Method (IRFM) as described by
\citet{casagrande10}. Each SSPP temperature estimate was re-scaled
to match the IRFM temperature estimate. This technique particularly
improves the temperature estimates for cool stars ($T_{\rm eff} < 5000$~K). 

\begin{figure*}[ht]
\centering
\plotone{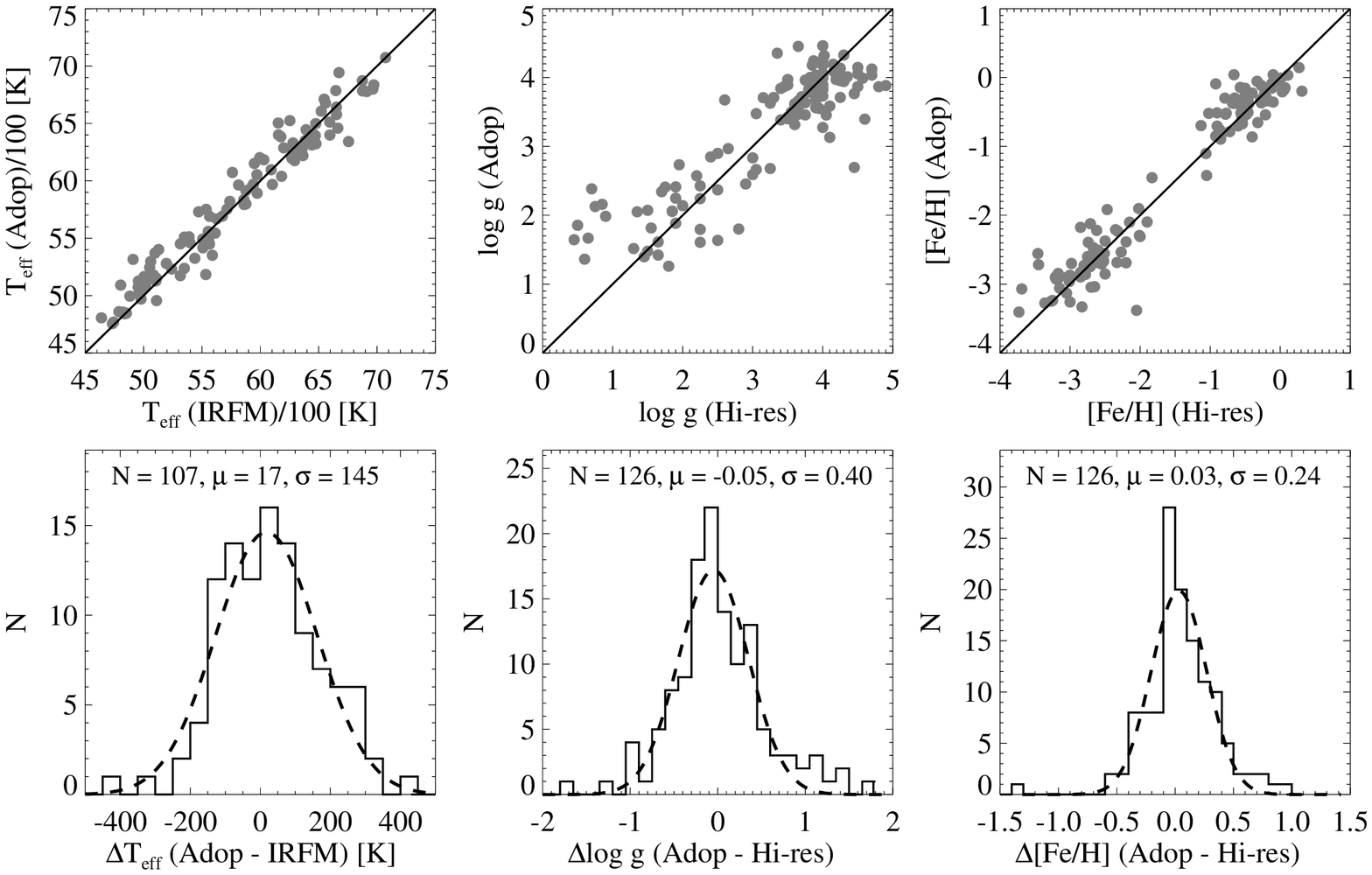}
\caption{Comparisons of $T_{\rm eff}$ (left panels), log $g$ (middle 
panels), and [Fe/H] (right panels) of the SSPP with the temperature from the
IRFM, and surface gravity and metallicity from analysis of
high-resolution spectra of 126 stars. The
symbols $\mu$ and $\sigma$ are the mean and standard deviation from a Gaussian
fit to the sample. `Adop' is the final adopted value in the SSPP; `Hi-res'
refers to the high-resolution analysis. As was the case for DR8, the DR9 SSPP gravity
value still over-estimates log $g$ by up to 1.0~dex for cool giants. 
There are only 107 stars available for the temperature
comparison, as $JHK$ photometry, needed to derive the IRFM
temperature, was unavailable for 19 stars. 
}
\label{fig:hires}
\end{figure*}

Figure \ref{fig:hires} shows the results of the comparisons of the SSPP
parameters with the IRFM for temperature, and the high-resolution analysis for
gravity and metallicity. Implementation of a grid of synthetic spectra with
microturbulences that vary appropriately with surface gravity also yields
improved estimates of metallicity for metal-rich stars ([Fe/H] $>-0.5$). 

A parameter comparison from a sample of about 9,000 multiply-observed stellar
spectra in SEGUE provides the basis for an estimate of the internal uncertainties of
the SSPP -- $\sim$50~K for $T_{\rm eff}$, $\sim$0.12~dex for log $g$, and
$\sim$0.10~dex for [Fe/H] for a typical G-type dwarf or redder 
stars in the color range of $0.4 < g-r < 1.3$ with S/N per pixel = 30. These
errors increase to $\sim$80~K, 0.30~dex, and 0.25~dex for $T_{\rm eff}$, log
$g$, and [Fe/H], respectively, for stars with $-0.3 < g-r < 0.2$, [Fe/H] $< -2.0$, and
S/N $< 15$.

A comparison with the DR8 parameters for stars from SEGUE-1 indicates
that the DR9 
average $T_{\rm eff}$ is higher by $\sim$60~K, the DR9 log $g$ is lower by
$\sim$0.2~dex, and the metallicity does not change significantly,
although these values vary with spectral type and spectral S/N.

These new SSPP results are made available for all stars in \mbox{SDSS-I/II},
including those of SEGUE-1 \citep{yanny09}, and the SEGUE-2 stars in
\mbox{SDSS-III}.   SSPP measurements are not currently available for the stars observed as
part of BOSS, although we plan to include that in future data
releases.

\section{Data Distribution}
\label{sec:distribution}

All Data Release 9 data are available through data access tools linked
from the DR9 web site.\footnote{\url{http://www.sdss3.org/dr9}}  
The data are stored both as flat files in the Science 
Archive Server (SAS),\footnote{The Science Archive Server (SAS) is the \mbox{SDSS-III} equivalent of the SDSS-I/II Data Archive Server (DAS).}
and as a searchable database in the 
Catalog Archive Server (CAS). 
A number of
different interfaces are available, each designed to accomplish a
specific task: 
(1) Color images of regions of the sky in JPEG format (based on the
$g, r$, and $i$ images; see \citealt{lupton04}) can be viewed in a web
browser with the SkyServer Navigate tool; 
(2) FITS images can be
downloaded through the SAS; 
(3) Complete
catalog information (astrometry, photometry, etc.) of any imaging
object can be viewed through the SkyServer Explore tool; and
(4) FITS files of the spectra can be downloaded through the SAS.

In addition, a number of catalog search tools are available through
the SkyServer interface to the CAS, each 
of which returns catalog data for objects that match supplied
criteria.  For more advanced queries, a powerful and flexible catalog
search website called ``CasJobs'' allows users to create their own
personalized data sets and 
then to modify or graph their data. 

The DR9 web site also features data access tutorials, a glossary of
SDSS terms, and detailed documentation about algorithms used to
process the imaging and spectroscopic data and select 
spectroscopic targets.

Imaging and spectroscopic data from all prior data releases are
also available through 
DR9 data access tools.  

\section{Conclusions}
\label{sec:conclusions}

The \mbox{SDSS-III} Data Release 9 presents the first data from the
BOSS survey, with $\sim$102,000 new quasar spectra, $\sim$91,000 new stellar
spectra and $\sim$536,000 new galaxy spectra.  The
astrometry has been improved since DR8, and the stellar
properties for SEGUE-I/II and \mbox{SDSS-I/II} stars have been updated. 

These data are already sufficient for cosmological analyses of large-scale structure, investigations of the structure of the Milky Way, measurements of quasar physics, clustering, and demographics, and countless other science investigations.
We invite the larger scientific community to investigate and explore this new data set.

The \mbox{SDSS-III} project will present two more public data releases.  
DR10, in summer 2013, will include
the first data from the APOGEE survey and another year of BOSS data.
DR11 will be an internal release only, as a public release would
occur only six months before 
the final public data
release for \mbox{SDSS-III}, DR12, which will be released in December 2014 and will contain
all of the data taken during the six years of the project. 

\acknowledgments
Funding for \mbox{SDSS-III} has been provided by the Alfred P. Sloan Foundation, the Participating Institutions, the National Science Foundation, and the U.S. Department of Energy Office of Science. The \mbox{SDSS-III} web site is http://www.sdss3.org/.

\mbox{SDSS-III} is managed by the Astrophysical Research Consortium for the Participating Institutions of the \mbox{SDSS-III} Collaboration including the University of Arizona, the Brazilian Participation Group, Brookhaven National Laboratory, University of Cambridge, Carnegie Mellon University, University of Florida, the French Participation Group, the German Participation Group, Harvard University, the Instituto de Astrofisica de Canarias, the Michigan State/Notre Dame/JINA Participation Group, Johns Hopkins University, Lawrence Berkeley National Laboratory, Max Planck Institute for Astrophysics, Max Planck Institute for Extraterrestrial Physics, New Mexico State University, New York University, Ohio State University, Pennsylvania State University, University of Portsmouth, Princeton University, the Spanish Participation Group, University of Tokyo, University of Utah, Vanderbilt University, University of Virginia, University of Washington, and Yale University.

\bibliographystyle{astroads}
\bibliography{refs}

\begin{thebibliography}{60}
\expandafter\ifx\csname natexlab\endcsname\relax\def\natexlab#1{#1}\fi
\expandafter\ifx\csname href\endcsname\relax
  \def\href#1#2{}\fi
\expandafter\ifx\csname urllinklabel\endcsname\relax
  \def\urllinklabel{[LINK]}\fi
\expandafter\ifx\csname adsurllinklabel\endcsname\relax
  \def\adsurllinklabel{[ADS]}\fi

\bibitem[{{Abazajian} {et~al.}(2009){Abazajian}, {Adelman-McCarthy},
  {Ag{\"u}eros}, {Allam}, {Allende Prieto}, {An}, {Anderson}, {Anderson},
  {Annis}, {Bahcall}, {et~al.}}]{DR7}
{Abazajian}, K.~N., et~al.\ 2009, \apjs, 182, 543

\bibitem[{{Adelman-McCarthy} {et~al.}(2008){Adelman-McCarthy}, {Ag{\"u}eros},
  {Allam}, {Allende Prieto}, {Anderson}, {Anderson}, {Annis}, {Bahcall},
  {Bailer-Jones}, {Baldry}, {Barentine}, {Bassett}, {Becker}, {Beers}, {Bell},
  {Berlind}, {Bernardi}, {Blanton}, {Bochanski}, {Boroski}, {Brinchmann},
  {Brinkmann}, {Brunner}, {Budav{\'a}ri}, {Carliles}, {Carr}, {Castander},
  {Cinabro}, {Cool}, {Covey}, {Csabai}, {Cunha}, {Davenport}, {Dilday}, {Doi},
  {Eisenstein}, {Evans}, {Fan}, {Finkbeiner}, {Friedman}, {Frieman},
  {Fukugita}, {G{\"a}nsicke}, {Gates}, {Gillespie}, {Glazebrook}, {Gray},
  {Grebel}, {Gunn}, {Gurbani}, {Hall}, {Harding}, {Harvanek}, {Hawley},
  {Hayes}, {Heckman}, {Hendry}, {Hindsley}, {Hirata}, {Hogan}, {Hogg}, {Hyde},
  {Ichikawa}, {Ivezi{\'c}}, {Jester}, {Johnson}, {Jorgensen}, {Juri{\'c}},
  {Kent}, {Kessler}, {Kleinman}, {Knapp}, {Kron}, {Krzesinski}, {Kuropatkin},
  {Lamb}, {Lampeitl}, {Lebedeva}, {Lee}, {Leger}, {L{\'e}pine}, {Lima}, {Lin},
  {Long}, {Loomis}, {Loveday}, {Lupton}, {Malanushenko}, {Malanushenko},
  {Mandelbaum}, {Margon}, {Marriner}, {Mart{\'{\i}}nez-Delgado}, {Matsubara},
  {McGehee}, {McKay}, {Meiksin}, {Morrison}, {Munn}, {Nakajima}, {Neilsen},
  {Newberg}, {Nichol}, {Nicinski}, {Nieto-Santisteban}, {Nitta}, {Okamura},
  {Owen}, {Oyaizu}, {Padmanabhan}, {Pan}, {Park}, {Peoples}, {Pier}, {Pope},
  {Purger}, {Raddick}, {Re Fiorentin}, {Richards}, {Richmond}, {Riess}, {Rix},
  {Rockosi}, {Sako}, {Schlegel}, {Schneider}, {Schreiber}, {Schwope}, {Seljak},
  {Sesar}, {Sheldon}, {Shimasaku}, {Sivarani}, {Smith}, {Snedden}, {Steinmetz},
  {Strauss}, {SubbaRao}, {Suto}, {Szalay}, {Szapudi}, {Szkody}, {Tegmark},
  {Thakar}, {Tremonti}, {Tucker}, {Uomoto}, {Vanden Berk}, {Vandenberg},
  {Vidrih}, {Vogeley}, {Voges}, {Vogt}, {Wadadekar}, {Weinberg}, {West},
  {White}, {Wilhite}, {Yanny}, {Yocum}, {York}, {Zehavi}, \& {Zucker}}]{DR6}
{Adelman-McCarthy}, J.~K., et~al.\ 2008, \apjs, 175, 297

\bibitem[{{Aihara} {et~al.}(2011{\natexlab{a}}){Aihara}, {Allende Prieto},
  {An}, {Anderson}, {Aubourg}, {Balbinot}, {Beers}, {Berlind}, {Bickerton},
  {Bizyaev}, {Blanton}, {Bochanski}, {Bolton}, {Bovy}, {Brandt}, {Brinkmann},
  {Brown}, {Brownstein}, {Busca}, {Campbell}, {Carr}, {Chen}, {Chiappini},
  {Comparat}, {Connolly}, {Cortes}, {Croft}, {Cuesta}, {da Costa}, {Davenport},
  {Dawson}, {Dhital}, {Ealet}, {Ebelke}, {Edmondson}, {Eisenstein},
  {Escoffier}, {Esposito}, {Evans}, {Fan}, {Femen{\'{\i}}a Castell{\'a}},
  {Font-Ribera}, {Frinchaboy}, {Ge}, {Gillespie}, {Gilmore}, {Gonz{\'a}lez
  Hern{\'a}ndez}, {Gott}, {Gould}, {Grebel}, {Gunn}, {Hamilton}, {Harding},
  {Harris}, {Hawley}, {Hearty}, {Ho}, {Hogg}, {Holtzman}, {Honscheid}, {Inada},
  {Ivans}, {Jiang}, {Johnson}, {Jordan}, {Jordan}, {Kazin}, {Kirkby}, {Klaene},
  {Knapp}, {Kneib}, {Kochanek}, {Koesterke}, {Kollmeier}, {Kron}, {Lampeitl},
  {Lang}, {Le Goff}, {Lee}, {Lin}, {Long}, {Loomis}, {Lucatello}, {Lundgren},
  {Lupton}, {Ma}, {MacDonald}, {Mahadevan}, {Maia}, {Makler}, {Malanushenko},
  {Malanushenko}, {Mandelbaum}, {Maraston}, {Margala}, {Masters}, {McBride},
  {McGehee}, {McGreer}, {M{\'e}nard}, {Miralda-Escud{\'e}}, {Morrison},
  {Mullally}, {Muna}, {Munn}, {Murayama}, {Myers}, {Naugle}, {Neto}, {Nguyen},
  {Nichol}, {O'Connell}, {Ogando}, {Olmstead}, {Oravetz}, {Padmanabhan},
  {Palanque-Delabrouille}, {Pan}, {Pandey}, {P{\^a}ris}, {Percival},
  {Petitjean}, {Pfaffenberger}, {Pforr}, {Phleps}, {Pichon}, {Pieri}, {Prada},
  {Price-Whelan}, {Raddick}, {Ramos}, {Reyl{\'e}}, {Rich}, {Richards}, {Rix},
  {Robin}, {Rocha-Pinto}, {Rockosi}, {Roe}, {Rollinde}, {Ross}, {Ross},
  {Rossetto}, {S{\'a}nchez}, {Sayres}, {Schlegel}, {Schlesinger}, {Schmidt},
  {Schneider}, {Sheldon}, {Shu}, {Simmerer}, {Simmons}, {Sivarani}, {Snedden},
  {Sobeck}, {Steinmetz}, {Strauss}, {Szalay}, {Tanaka}, {Thakar}, {Thomas},
  {Tinker}, {Tofflemire}, {Tojeiro}, {Tremonti}, {Vandenberg}, {Vargas
  Maga{\~n}a}, {Verde}, {Vogt}, {Wake}, {Wang}, {Weaver}, {Weinberg}, {White},
  {White}, {Yanny}, {Yasuda}, {Yeche}, \& {Zehavi}}]{DR8}
{Aihara}, H., et~al.\ 2011{\natexlab{a}}, \apjs,  193, 29

\bibitem[{{Aihara} {et~al.}(2011{\natexlab{b}}){Aihara}, {Allende Prieto},
  {An}, {Anderson}, {Aubourg}, {Balbinot}, {Beers}, {Berlind}, {Bickerton},
  {Bizyaev}, {Blanton}, {Bochanski}, {Bolton}, {Bovy}, {Brandt}, {Brinkmann},
  {Brown}, {Brownstein}, {Busca}, {Campbell}, {Carr}, {Chen}, {Chiappini},
  {Comparat}, {Connolly}, {Cortes}, {Croft}, {Cuesta}, {da Costa}, {Davenport},
  {Dawson}, {Dhital}, {Ealet}, {Ebelke}, {Edmondson}, {Eisenstein},
  {Escoffier}, {Esposito}, {Evans}, {Fan}, {Femen{\'{\i}}a Castell{\'a}},
  {Font-Ribera}, {Frinchaboy}, {Ge}, {Gillespie}, {Gilmore}, {Gonz{\'a}lez
  Hern{\'a}ndez}, {Gott}, {Gould}, {Grebel}, {Gunn}, {Hamilton}, {Harding},
  {Harris}, {Hawley}, {Hearty}, {Ho}, {Hogg}, {Holtzman}, {Honscheid}, {Inada},
  {Ivans}, {Jiang}, {Johnson}, {Jordan}, {Jordan}, {Kazin}, {Kirkby}, {Klaene},
  {Knapp}, {Kneib}, {Kochanek}, {Koesterke}, {Kollmeier}, {Kron}, {Lampeitl},
  {Lang}, {Le Goff}, {Lee}, {Lin}, {Long}, {Loomis}, {Lucatello}, {Lundgren},
  {Lupton}, {Ma}, {MacDonald}, {Mahadevan}, {Maia}, {Makler}, {Malanushenko},
  {Malanushenko}, {Mandelbaum}, {Maraston}, {Margala}, {Masters}, {McBride},
  {McGehee}, {McGreer}, {M{\'e}nard}, {Miralda-Escud{\'e}}, {Morrison},
  {Mullally}, {Muna}, {Munn}, {Murayama}, {Myers}, {Naugle}, {Neto}, {Nguyen},
  {Nichol}, {O'Connell}, {Ogando}, {Olmstead}, {Oravetz}, {Padmanabhan},
  {Palanque-Delabrouille}, {Pan}, {Pandey}, {P{\^a}ris}, {Percival},
  {Petitjean}, {Pfaffenberger}, {Pforr}, {Phleps}, {Pichon}, {Pieri}, {Prada},
  {Price-Whelan}, {Raddick}, {Ramos}, {Reyl{\'e}}, {Rich}, {Richards}, {Rix},
  {Robin}, {Rocha-Pinto}, {Rockosi}, {Roe}, {Rollinde}, {Ross}, {Ross},
  {Rossetto}, {S{\'a}nchez}, {Sayres}, {Schlegel}, {Schlesinger}, {Schmidt},
  {Schneider}, {Sheldon}, {Shu}, {Simmerer}, {Simmons}, {Sivarani}, {Snedden},
  {Sobeck}, {Steinmetz}, {Strauss}, {Szalay}, {Tanaka}, {Thakar}, {Thomas},
  {Tinker}, {Tofflemire}, {Tojeiro}, {Tremonti}, {Vandenberg}, {Vargas
  Maga{\~n}a}, {Verde}, {Vogt}, {Wake}, {Wang}, {Weaver}, {Weinberg}, {White},
  {White}, {Yanny}, {Yasuda}, {Yeche}, \& {Zehavi}}]{DR8_erratum}
---. 2011{\natexlab{b}}, \apjs, 195, 26


\bibitem[{{Allende Prieto} {et~al.}(2008){Allende Prieto}, {Sivarani}, {Beers},
  {Lee}, {Koesterke}, {Shetrone}, {Sneden}, {Lambert}, {Wilhelm}, {Rockosi},
  {Lai}, {Yanny}, {Ivans}, {Johnson}, {Aoki}, {Bailer-Jones}, \& {Re
  Fiorentin}}]{allendeprieto08}
{Allende Prieto}, C., et~al.\ 2008, \aj,  136, 2070

\bibitem[{{Anderson} {et~al.}(2012){Anderson}, {Aubourg}, {Bailey}, {Bizyaev},
  {Blanton}, {Bolton}, {Brinkmann}, {Brownstein}, {Burden}, {Cuesta}, {da
  Costa}, {Dawson}, {de Putter}, {Eisenstein}, {Gunn}, {Guo}, {Hamilton},
  {Harding}, {Ho}, {Honscheid}, {Kazin}, {Kirkby}, {Kneib}, {Labatie},
  {Loomis}, {Lupton}, {Malanushenko}, {Malanushenko}, {Mandelbaum}, {Manera},
  {Maraston}, {McBride}, {Mehta}, {Mena}, {Montesano}, {Muna}, {Nichol},
  {Nuza}, {Olmstead}, {Oravetz}, {Padmanabhan}, {Palanque-Delabrouille}, {Pan},
  {Parejko}, {Paris}, {Percival}, {Petitjean}, {Prada}, {Reid}, {Roe}, {Ross},
  {Ross}, {Samushia}, {Sanchez}, {Schneider}, {Scoccola}, {Seo}, {Sheldon},
  {Simmons}, {Skibba}, {Strauss}, {Swanson}, {Thomas}, {Tinker}, {Tojeiro},
  {Vargas Magana}, {Verde}, {Wagner}, {Wake}, {Weaver}, {Weinberg}, {White},
  {Xu}, {Yeche}, {Zehavi}, \& {Zhao}}]{anderson12}
{Anderson}, L., et~al.\ 2012, ArXiv  e-prints

\bibitem[{{Bolton} {et~al.}(2012)}]{bolton12}
{Bolton}, A. {et~al.} 2012, {{\it submitted to \aj}}


\bibitem[{{Bond} {et~al.}(2010){Bond}, {Ivezi{\'c}}, {Sesar}, {Juri{\'c}},
  {Munn}, {Kowalski}, {Loebman}, {Ro{\v s}kar}, {Beers}, {Dalcanton},
  {Rockosi}, {Yanny}, {Newberg}, {Allende Prieto}, {Wilhelm}, {Lee},
  {Sivarani}, {Majewski}, {Norris}, {Bailer-Jones}, {Re Fiorentin}, {Schlegel},
  {Uomoto}, {Lupton}, {Knapp}, {Gunn}, {Covey}, {Allyn Smith}, {Miknaitis},
  {Doi}, {Tanaka}, {Fukugita}, {Kent}, {Finkbeiner}, {Quinn}, {Hawley},
  {Anderson}, {Kiuchi}, {Chen}, {Bushong}, {Sohi}, {Haggard}, {Kimball},
  {McGurk}, {Barentine}, {Brewington}, {Harvanek}, {Kleinman}, {Krzesinski},
  {Long}, {Nitta}, {Snedden}, {Lee}, {Pier}, {Harris}, {Brinkmann}, \&
  {Schneider}}]{bond10a}
{Bond}, N.~A., et~al.\ 2010, \apj, 716, 1

\bibitem[{{Bovy} {et~al.}(2011){Bovy}, {Hennawi}, {Hogg}, {Myers},
  {Kirkpatrick}, {Schlegel}, {Ross}, {Sheldon}, {McGreer}, {Schneider}, \&
  {Weaver}}]{bovy11}
{Bovy}, J., et~al.\ 2011, \apj, 729, 141

\bibitem[{{Brinchmann} {et~al.}(2004){Brinchmann}, {Charlot}, {White},
  {Tremonti}, {Kauffmann}, {Heckman}, \& {Brinkmann}}]{2004MNRAS.351.1151B}
{Brinchmann}, J., {Charlot}, S., {White}, S.~D.~M., {Tremonti}, C.,
  {Kauffmann}, G., {Heckman}, T., \& {Brinkmann}, J. 2004, \mnras, 351, 1151


\bibitem[{{Bruzual} \& {Charlot}(2003)}]{2003MNRAS.344.1000B}
{Bruzual}, G. \& {Charlot}, S. 2003, \mnras, 344, 1000


\bibitem[{{Cappellari} \& {Emsellem}(2004)}]{2004PASP..116..138C}
{Cappellari}, M. \& {Emsellem}, E. 2004, \pasp, 116, 138


\bibitem[{{Casagrande} {et~al.}(2010){Casagrande}, {Ram{\'{\i}}rez},
  {Mel{\'e}ndez}, {Bessell}, \& {Asplund}}]{casagrande10}
{Casagrande}, L., {Ram{\'{\i}}rez}, I., {Mel{\'e}ndez}, J., {Bessell}, M., \&
  {Asplund}, M. 2010, \aap, 512, A54


\bibitem[{{Chen} {et~al.}(2012){Chen}, {Kauffmann}, {Tremonti}, {White},
  {Heckman}, {Kova{\v c}}, {Bundy}, {Chisholm}, {Maraston}, {Schneider},
  {Bolton}, {Weaver}, \& {Brinkmann}}]{chen12}
{Chen}, Y.-M., et~al.\ 2012, \mnras, 421, 314

\bibitem[{{Cole} {et~al.}(2005){Cole}, {Percival}, {Peacock}, {Norberg},
  {Baugh}, {Frenk}, {Baldry}, {Bland-Hawthorn}, {Bridges}, {Cannon}, {Colless},
  {Collins}, {Couch}, {Cross}, {Dalton}, {Eke}, {De Propris}, {Driver},
  {Efstathiou}, {Ellis}, {Glazebrook}, {Jackson}, {Jenkins}, {Lahav}, {Lewis},
  {Lumsden}, {Maddox}, {Madgwick}, {Peterson}, {Sutherland}, \&
  {Taylor}}]{cole05}
{Cole}, S., et~al.\ 2005, \mnras, 362, 505

\bibitem[{{Dawson} {et~al.}(2012)}]{dawson12}
{Dawson}, K. {et~al.} 2012, {\it submitted to \aj}


\bibitem[{{Eisenstein} {et~al.}(2001){Eisenstein}, {Annis}, {Gunn}, {Szalay},
  {Connolly}, {Nichol}, {Bahcall}, {Bernardi}, {Burles}, {Castander},
  {Fukugita}, {Hogg}, {Ivezi{\'c}}, {Knapp}, {Lupton}, {Narayanan}, {Postman},
  {Reichart}, {Richmond}, {Schneider}, {Schlegel}, {Strauss}, {SubbaRao},
  {Tucker}, {Vanden Berk}, {Vogeley}, {Weinberg}, \& {Yanny}}]{eisenstein01}
{Eisenstein}, D.~J., et~al.\ 2001, \aj, 122, 2267

\bibitem[{{Eisenstein} {et~al.}(2011){Eisenstein}, {Weinberg}, {Agol},
  {Aihara}, {Allende Prieto}, {Anderson}, {Arns}, {Aubourg}, {Bailey},
  {Balbinot}, {et~al.}}]{eisenstein11}
{Eisenstein}, D.~J., et~al.\ 2011, \aj, 142, 72

\bibitem[{{Eisenstein} {et~al.}(2005){Eisenstein}, {Zehavi}, {Hogg},
  {Scoccimarro}, {Blanton}, {Nichol}, {Scranton}, {Seo}, {Tegmark}, {Zheng},
  {Anderson}, {Annis}, {Bahcall}, {Brinkmann}, {Burles}, {Castander},
  {Connolly}, {Csabai}, {Doi}, {Fukugita}, {Frieman}, {Glazebrook}, {Gunn},
  {Hendry}, {Hennessy}, {Ivezi{\'c}}, {Kent}, {Knapp}, {Lin}, {Loh}, {Lupton},
  {Margon}, {McKay}, {Meiksin}, {Munn}, {Pope}, {Richmond}, {Schlegel},
  {Schneider}, {Shimasaku}, {Stoughton}, {Strauss}, {SubbaRao}, {Szalay},
  {Szapudi}, {Tucker}, {Yanny}, \& {York}}]{eisenstein05}
{Eisenstein}, D.~J., et~al.\ 2005, \apj, 633, 560

\bibitem[{{Fan}(1999)}]{fan99}
{Fan}, X. 1999, \aj, 117, 2528


\bibitem[{{Fukugita} {et~al.}(1996){Fukugita}, {Ichikawa}, {Gunn}, {Doi},
  {Shimasaku}, \& {Schneider}}]{fukugita96}
{Fukugita}, M., {Ichikawa}, T., {Gunn}, J.~E., {Doi}, M., {Shimasaku}, K., \&
  {Schneider}, D.~P. 1996, \aj, 111, 1748


\bibitem[{{Gunn} {et~al.}(2006){Gunn}, {Siegmund}, {Mannery}, {Owen}, {Hull},
  {Leger}, {Carey}, {Knapp}, {York}, {Boroski}, {Kent}, {Lupton}, {Rockosi},
  {Evans}, {Waddell}, {Anderson}, {Annis}, {Barentine}, {Bartoszek}, {Bastian},
  {Bracker}, {Brewington}, {Briegel}, {Brinkmann}, {Brown}, {Carr},
  {Czarapata}, {Drennan}, {Dombeck}, {Federwitz}, {Gillespie}, {Gonzales},
  {Hansen}, {Harvanek}, {Hayes}, {Jordan}, {Kinney}, {Klaene}, {Kleinman},
  {Kron}, {Kresinski}, {Lee}, {Limmongkol}, {Lindenmeyer}, {Long}, {Loomis},
  {McGehee}, {Mantsch}, {Neilsen}, {Neswold}, {Newman}, {Nitta}, {Peoples},
  {Pier}, {Prieto}, {Prosapio}, {Rivetta}, {Schneider}, {Snedden}, \&
  {Wang}}]{gunn06}
{Gunn}, J.~E., et~al.\ 2006, \aj,  131, 2332

\bibitem[{{Hewett} \& {Wild}(2010)}]{hewett10}
{Hewett}, P.~C. \& {Wild}, V. 2010, \mnras, 405, 2302


\bibitem[{{Kauffmann} {et~al.}(2003){Kauffmann}, {Heckman}, {White}, {Charlot},
  {Tremonti}, {Brinchmann}, {Bruzual}, {Peng}, {Seibert}, {Bernardi},
  {Blanton}, {Brinkmann}, {Castander}, {Cs{\'a}bai}, {Fukugita}, {Ivezic},
  {Munn}, {Nichol}, {Padmanabhan}, {Thakar}, {Weinberg}, \&
  {York}}]{2003MNRAS.341...33K}
{Kauffmann}, G., et~al.\ 2003, \mnras, 341, 33

\bibitem[{{Komatsu} {et~al.}(2011){Komatsu}, {Smith}, {Dunkley}, {Bennett},
  {Gold}, {Hinshaw}, {Jarosik}, {Larson}, {Nolta}, {Page}, {Spergel},
  {Halpern}, {Hill}, {Kogut}, {Limon}, {Meyer}, {Odegard}, {Tucker}, {Weiland},
  {Wollack}, \& {Wright}}]{komatsu11}
{Komatsu}, E., et~al.\ 2011, \apjs, 192, 18

\bibitem[{{Kroupa}(2001)}]{kroupa01}
{Kroupa}, P. 2001, \mnras, 322, 231


\bibitem[{{Lee} {et~al.}(2012)}]{lee12}
{Lee}, K.-G. {et~al.} 2012, {\it in prep}


\bibitem[{{Lee} {et~al.}(2008{\natexlab{a}}){Lee}, {Beers}, {Sivarani},
  {Allende Prieto}, {Koesterke}, {Wilhelm}, {Re Fiorentin}, {Bailer-Jones},
  {Norris}, {Rockosi}, {Yanny}, {Newberg}, {Covey}, {Zhang}, \& {Luo}}]{lee08a}
{Lee}, Y.~S., et~al.\ 2008{\natexlab{a}}, \aj, 136, 2022

\bibitem[{{Lee} {et~al.}(2008{\natexlab{b}}){Lee}, {Beers}, {Sivarani},
  {Johnson}, {An}, {Wilhelm}, {Allende Prieto}, {Koesterke}, {Re Fiorentin},
  {Bailer-Jones}, {Norris}, {Yanny}, {Rockosi}, {Newberg}, {Cudworth}, \&
  {Pan}}]{lee08b}
{Lee}, Y.~S., et~al.\ 2008{\natexlab{b}}, \aj,  136, 2050

\bibitem[{{Lupton} {et~al.}(2004){Lupton}, {Blanton}, {Fekete}, {Hogg},
  {O'Mullane}, {Szalay}, \& {Wherry}}]{lupton04}
{Lupton}, R., {Blanton}, M.~R., {Fekete}, G., {Hogg}, D.~W., {O'Mullane}, W.,
  {Szalay}, A., \& {Wherry}, N. 2004, \pasp, 116, 133


\bibitem[{{Maraston}(2005)}]{2005MNRAS.362..799M}
{Maraston}, C. 2005, \mnras, 362, 799


\bibitem[{{Maraston} {et~al.}(2012){Maraston}, {Pforr}, {Henriques}, {Thomas},
  {Wake}, {Brownstein}, {Capozzi}, {Bundy}, {Skibba}, {Beifiori}, {Nichol},
  {Edmondson}, {Schneider}, {Chen}, {Masters}, {Steele}, {Bolton}, {York},
  {Bizyaev}, {Brewington}, {Malanushenko}, {Malanushenko}, {Snedden},
  {Oravetz}, {Pan}, {Shelden}, \& {Simmons}}]{maraston12}
{Maraston}, C., {Pforr}, J., {Henriques}, B.~M., {Thomas}, D., {Wake}, D.,
  {Brownstein}, J.~R., {Capozzi}, D., {Bundy}, K., {Skibba}, R.~A., {Beifiori},
  A., {Nichol}, R.~C., {Edmondson}, E., {Schneider}, D.~P., {Chen}, Y.,
  {Masters}, K.~L., {Steele}, O., {Bolton}, A.~S., {York}, D.~G., {Bizyaev},
  D., {Brewington}, H., {Malanushenko}, E., {Malanushenko}, V., {Snedden}, S.,
  {Oravetz}, D., {Pan}, K., {Shelden}, A., \& {Simmons}, A. 2012, ArXiv
  e-prints, {\it submitted to MNRAS.}


\bibitem[{{Maraston} \& {Str{\"o}mb{\"a}ck}(2011)}]{2011MNRAS.418.2785M}
{Maraston}, C. \& {Str{\"o}mb{\"a}ck}, G. 2011, \mnras, 418, 2785


\bibitem[{{Maraston} {et~al.}(2009){Maraston}, {Str{\"o}mb{\"a}ck}, {Thomas},
  {Wake}, \& {Nichol}}]{2009MNRAS.394L.107M}
{Maraston}, C., {Str{\"o}mb{\"a}ck}, G., {Thomas}, D., {Wake}, D.~A., \&
  {Nichol}, R.~C. 2009, \mnras, 394, L107


\bibitem[{{McDonald} \& {Eisenstein}(2007)}]{mcdonald07}
{McDonald}, P. \& {Eisenstein}, D.~J. 2007, \prd, 76, 063009


\bibitem[{{McQuinn} \& {White}(2011)}]{mcquinn11}
{McQuinn}, M. \& {White}, M. 2011, \mnras, 415, 2257


\bibitem[{{Monet} {et~al.}(2003){Monet}, {Levine}, {Canzian}, {Ables}, {Bird},
  {Dahn}, {Guetter}, {Harris}, {Henden}, {Leggett}, {Levison}, {Luginbuhl},
  {Martini}, {Monet}, {Munn}, {Pier}, {Rhodes}, {Riepe}, {Sell}, {Stone},
  {Vrba}, {Walker}, {Westerhout}, {Brucato}, {Reid}, {Schoening}, {Hartley},
  {Read}, \& {Tritton}}]{monet03}
{Monet}, D.~G., et~al.\ 2003, \aj, 125, 984

\bibitem[{{Padmanabhan} {et~al.}(2012{\natexlab{a}}){Padmanabhan}, {Xu},
  {Eisenstein}, {Scalzo}, {Cuesta}, {Mehta}, \& {Kazin}}]{padmanabhan12a}
{Padmanabhan}, N., {Xu}, X., {Eisenstein}, D.~J., {Scalzo}, R., {Cuesta},
  A.~J., {Mehta}, K.~T., \& {Kazin}, E. 2012{\natexlab{a}}, ArXiv e-prints


\bibitem[{{Padmanabhan} {et~al.}(2012{\natexlab{b}})}]{padmanabhan12b}
{Padmanabhan}, N. {et~al.} 2012{\natexlab{b}}, {\it in prep}


\bibitem[{{Palanque-Delabrouille} {et~al.}(2011){Palanque-Delabrouille},
  {Yeche}, {Myers}, {Petitjean}, {Ross}, {Sheldon}, {Aubourg}, {Delubac}, {Le
  Goff}, {P{\^a}ris}, {Rich}, {Dawson}, {Schneider}, \& {Weaver}}]{palanque11}
{Palanque-Delabrouille}, N., et~al.\ 2011, \aap, 530, A122

\bibitem[{{P\^aris} {et~al.}(2012)}]{paris12}
{P\^aris}, I. {et~al.} 2012, {\it submitted to \aap}


\bibitem[{{Percival} {et~al.}(2010){Percival}, {Reid}, {Eisenstein}, {Bahcall},
  {Budavari}, {Frieman}, {Fukugita}, {Gunn}, {Ivezi{\'c}}, {Knapp}, {Kron},
  {Loveday}, {Lupton}, {McKay}, {Meiksin}, {Nichol}, {Pope}, {Schlegel},
  {Schneider}, {Spergel}, {Stoughton}, {Strauss}, {Szalay}, {Tegmark},
  {Vogeley}, {Weinberg}, {York}, \& {Zehavi}}]{percival10}
{Percival}, W.~J., et~al.\ 2010, \mnras, 401, 2148

\bibitem[{{Richards} {et~al.}(2002){Richards}, {Fan}, {Newberg}, {Strauss},
  {Vanden Berk}, {Schneider}, {Yanny}, {Boucher}, {Burles}, {Frieman}, {Gunn},
  {Hall}, {Ivezi{\'c}}, {Kent}, {Loveday}, {Lupton}, {Rockosi}, {Schlegel},
  {Stoughton}, {SubbaRao}, \& {York}}]{richards02}
{Richards}, G.~T., et~al.\ 2002, \aj, 123, 2945

\bibitem[{{Rockosi} {et~al.}(2012)}]{rockosi12}
{Rockosi}, C. {et~al.} 2012, {\it in prep}


\bibitem[{{Ross} {et~al.}(2012){Ross}, {Myers}, {Sheldon}, {Y{\`e}che},
  {Strauss}, {Bovy}, {Kirkpatrick}, {Richards}, {Aubourg}, {Blanton}, {Brandt},
  {Carithers}, {Croft}, {da Silva}, {Dawson}, {Eisenstein}, {Hennawi}, {Ho},
  {Hogg}, {Lee}, {Lundgren}, {McMahon}, {Miralda-Escud{\'e}},
  {Palanque-Delabrouille}, {P{\^a}ris}, {Petitjean}, {Pieri}, {Rich}, {Roe},
  {Schiminovich}, {Schlegel}, {Schneider}, {Slosar}, {Suzuki}, {Tinker},
  {Weinberg}, {Weyant}, {White}, \& {Wood-Vasey}}]{ross12}
{Ross}, N.~P., et~al.\ 2012, \apjs, 199, 3

\bibitem[{{Salpeter}(1955)}]{salpeter55}
{Salpeter}, E.~E. 1955, \apj, 121, 161


\bibitem[{{Sarzi} {et~al.}(2006){Sarzi}, {Falc{\'o}n-Barroso}, {Davies},
  {Bacon}, {Bureau}, {Cappellari}, {de Zeeuw}, {Emsellem}, {Fathi},
  {Krajnovi{\'c}}, {Kuntschner}, {McDermid}, \&
  {Peletier}}]{2006MNRAS.366.1151S}
{Sarzi}, M., et~al.\ 2006, \mnras, 366, 1151

\bibitem[{{Schlegel} {et~al.}(2012)}]{schlegel12}
{Schlegel}, D. {et~al.} 2012, {\it in prep}


\bibitem[{{Schneider} {et~al.}(2010){Schneider}, {Richards}, {Hall}, {Strauss},
  {Anderson}, {Boroson}, {Ross}, {Shen}, {Brandt}, {Fan}, {Inada}, {Jester},
  {Knapp}, {Krawczyk}, {Thakar}, {Vanden Berk}, {Voges}, {Yanny}, {York},
  {Bahcall}, {Bizyaev}, {Blanton}, {Brewington}, {Brinkmann}, {Eisenstein},
  {Frieman}, {Fukugita}, {Gray}, {Gunn}, {Hibon}, {Ivezi{\'c}}, {Kent}, {Kron},
  {Lee}, {Lupton}, {Malanushenko}, {Malanushenko}, {Oravetz}, {Pan}, {Pier},
  {Price}, {Saxe}, {Schlegel}, {Simmons}, {Snedden}, {SubbaRao}, {Szalay}, \&
  {Weinberg}}]{DR7Q}
{Schneider}, D.~P., et~al.\ 2010, \aj, 139, 2360

\bibitem[{{Slosar} {et~al.}(2011){Slosar}, {Font-Ribera}, {Pieri}, {Rich}, {Le
  Goff}, {Aubourg}, {Brinkmann}, {Busca}, {Carithers}, {Charlassier},
  {Cort{\^e}s}, {Croft}, {Dawson}, {Eisenstein}, {Hamilton}, {Ho}, {Lee},
  {Lupton}, {McDonald}, {Medolin}, {Muna}, {Miralda-Escud{\'e}}, {Myers},
  {Nichol}, {Palanque-Delabrouille}, {P{\^a}ris}, {Petitjean}, {Pi{\v s}kur},
  {Rollinde}, {Ross}, {Schlegel}, {Schneider}, {Sheldon}, {Weaver}, {Weinberg},
  {Yeche}, \& {York}}]{slosar11}
{Slosar}, A., et~al.\ 2011, \jcap, 9, 1

\bibitem[{{Smee} {et~al.}(2012)}]{smee12}
{Smee}, S. {et~al.} 2012, {\it in prep}


\bibitem[{{Smolinski} {et~al.}(2011){Smolinski}, {Lee}, {Beers}, {An},
  {Bickerton}, {Johnson}, {Loomis}, {Rockosi}, {Sivarani}, \&
  {Yanny}}]{smolinski11}
{Smolinski}, J.~P., et~al.\ 2011, \aj, 141, 89

\bibitem[{{Strauss} {et~al.}(2002){Strauss}, {Weinberg}, {Lupton}, {Narayanan},
  {Annis}, {Bernardi}, {Blanton}, {Burles}, {Connolly}, {Dalcanton}, {Doi},
  {Eisenstein}, {Frieman}, {Fukugita}, {Gunn}, {Ivezi{\'c}}, {Kent}, {Kim},
  {Knapp}, {Kron}, {Munn}, {Newberg}, {Nichol}, {Okamura}, {Quinn}, {Richmond},
  {Schlegel}, {Shimasaku}, {SubbaRao}, {Szalay}, {Vanden Berk}, {Vogeley},
  {Yanny}, {Yasuda}, {York}, \& {Zehavi}}]{strauss02}
{Strauss}, M.~A., et~al.\ 2002, \aj, 124, 1810

\bibitem[{{Tegmark} {et~al.}(2006){Tegmark}, {Eisenstein}, {Strauss},
  {Weinberg}, {Blanton}, {Frieman}, {Fukugita}, {Gunn}, {Hamilton}, {Knapp},
  {Nichol}, {Ostriker}, {Padmanabhan}, {Percival}, {Schlegel}, {Schneider},
  {Scoccimarro}, {Seljak}, {Seo}, {Swanson}, {Szalay}, {Vogeley}, {Yoo},
  {Zehavi}, {Abazajian}, {Anderson}, {Annis}, {Bahcall}, {Bassett}, {Berlind},
  {Brinkmann}, {Budavari}, {Castander}, {Connolly}, {Csabai}, {Doi},
  {Finkbeiner}, {Gillespie}, {Glazebrook}, {Hennessy}, {Hogg}, {Ivezi{\'c}},
  {Jain}, {Johnston}, {Kent}, {Lamb}, {Lee}, {Lin}, {Loveday}, {Lupton},
  {Munn}, {Pan}, {Park}, {Peoples}, {Pier}, {Pope}, {Richmond}, {Rockosi},
  {Scranton}, {Sheth}, {Stebbins}, {Stoughton}, {Szapudi}, {Tucker}, {vanden
  Berk}, {Yanny}, \& {York}}]{tegmark06}
{Tegmark}, M., et~al.\ 2006, \prd, 74, 123507

\bibitem[{{Thomas} {et~al.}(2012){Thomas}, {Steele}, {Maraston}, {Johansson},
  {Beifiori}, {Pforr}, {Strombaeck}, {Tremonti}, {Wake}, {Bizyaev}, {Bolton},
  {Brewington}, {Brownstein}, {Comparat}, {Kneib}, {Malanushenko},
  {Malanushenko}, {Oravetz}, {Pan}, {Parejko}, {Schneider}, {Shelden},
  {Simmons}, {Snedden}, {Tanaka}, {Weaver}, \& {Yan}}]{thomas12}
{Thomas}, D., {Steele}, O., {Maraston}, C., {Johansson}, J., {Beifiori}, A.,
  {Pforr}, J., {Strombaeck}, G., {Tremonti}, C.~A., {Wake}, D., {Bizyaev}, D.,
  {Bolton}, A., {Brewington}, H., {Brownstein}, J.~R., {Comparat}, J., {Kneib},
  J.~P., {Malanushenko}, E., {Malanushenko}, V., {Oravetz}, D., {Pan}, K.,
  {Parejko}, J.~K., {Schneider}, D.~P., {Shelden}, A., {Simmons}, A.,
  {Snedden}, S., {Tanaka}, M., {Weaver}, B.~A., \& {Yan}, R. 2012, ArXiv
  e-prints, {\it submitted to MNRAS.}


\bibitem[{{Tremonti} {et~al.}(2004){Tremonti}, {Heckman}, {Kauffmann},
  {Brinchmann}, {Charlot}, {White}, {Seibert}, {Peng}, {Schlegel}, {Uomoto},
  {Fukugita}, \& {Brinkmann}}]{2004ApJ...613..898T}
{Tremonti}, C.~A., et~al.\ 2004, \apj, 613, 898

\bibitem[{{Weinberg} {et~al.}(2012){Weinberg}, {Mortonson}, {Eisenstein},
  {Hirata}, {Riess}, \& {Rozo}}]{weinberg12}
{Weinberg}, D.~H., {Mortonson}, M.~J., {Eisenstein}, D.~J., {Hirata}, C.,
  {Riess}, A.~G., \& {Rozo}, E. 2012, ArXiv e-prints


\bibitem[{{Yanny} {et~al.}(2009){Yanny}, {Rockosi}, {Newberg}, {Knapp},
  {Adelman-McCarthy}, {Alcorn}, {Allam}, {Allende Prieto}, {An}, {Anderson},
  {Anderson}, {Bailer-Jones}, {Bastian}, {Beers}, {Bell}, {Belokurov},
  {Bizyaev}, {Blythe}, {Bochanski}, {Boroski}, {Brinchmann}, {Brinkmann},
  {Brewington}, {Carey}, {Cudworth}, {Evans}, {Evans}, {Gates}, {G{\"a}nsicke},
  {Gillespie}, {Gilmore}, {Nebot Gomez-Moran}, {Grebel}, {Greenwell}, {Gunn},
  {Jordan}, {Jordan}, {Harding}, {Harris}, {Hendry}, {Holder}, {Ivans},
  {Ivezi{\v c}}, {Jester}, {Johnson}, {Kent}, {Kleinman}, {Kniazev},
  {Krzesinski}, {Kron}, {Kuropatkin}, {Lebedeva}, {Lee}, {French Leger},
  {L{\'e}pine}, {Levine}, {Lin}, {Long}, {Loomis}, {Lupton}, {Malanushenko},
  {Malanushenko}, {Margon}, {Martinez-Delgado}, {McGehee}, {Monet}, {Morrison},
  {Munn}, {Neilsen}, {Nitta}, {Norris}, {Oravetz}, {Owen}, {Padmanabhan},
  {Pan}, {Peterson}, {Pier}, {Platson}, {Re Fiorentin}, {Richards}, {Rix},
  {Schlegel}, {Schneider}, {Schreiber}, {Schwope}, {Sibley}, {Simmons},
  {Snedden}, {Allyn Smith}, {Stark}, {Stauffer}, {Steinmetz}, {Stoughton},
  {SubbaRao}, {Szalay}, {Szkody}, {Thakar}, {Thirupathi}, {Tucker}, {Uomoto},
  {Vanden Berk}, {Vidrih}, {Wadadekar}, {Watters}, {Wilhelm}, {Wyse}, {Yarger},
  \& {Zucker}}]{yanny09}
{Yanny}, B., et~al.\ 2009, \aj, 137, 4377

\bibitem[{{York} {et~al.}(2000){York}, {Adelman}, {Anderson}, {Anderson},
  {Annis}, {Bahcall}, {Bakken}, {Barkhouser}, {Bastian}, {Berman}, {Boroski},
  {Bracker}, {Briegel}, {Briggs}, {Brinkmann}, {Brunner}, {Burles}, {Carey},
  {Carr}, {Castander}, {Chen}, {Colestock}, {Connolly}, {Crocker}, {Csabai},
  {Czarapata}, {Davis}, {Doi}, {Dombeck}, {Eisenstein}, {Ellman}, {Elms},
  {Evans}, {Fan}, {Federwitz}, {Fiscelli}, {Friedman}, {Frieman}, {Fukugita},
  {Gillespie}, {Gunn}, {Gurbani}, {de Haas}, {Haldeman}, {Harris}, {Hayes},
  {Heckman}, {Hennessy}, {Hindsley}, {Holm}, {Holmgren}, {Huang}, {Hull},
  {Husby}, {Ichikawa}, {Ichikawa}, {Ivezi{\'c}}, {Kent}, {Kim}, {Kinney},
  {Klaene}, {Kleinman}, {Kleinman}, {Knapp}, {Korienek}, {Kron}, {Kunszt},
  {Lamb}, {Lee}, {Leger}, {Limmongkol}, {Lindenmeyer}, {Long}, {Loomis},
  {Loveday}, {Lucinio}, {Lupton}, {MacKinnon}, {Mannery}, {Mantsch}, {Margon},
  {McGehee}, {McKay}, {Meiksin}, {Merelli}, {Monet}, {Munn}, {Narayanan},
  {Nash}, {Neilsen}, {Neswold}, {Newberg}, {Nichol}, {Nicinski}, {Nonino},
  {Okada}, {Okamura}, {Ostriker}, {Owen}, {Pauls}, {Peoples}, {Peterson},
  {Petravick}, {Pier}, {Pope}, {Pordes}, {Prosapio}, {Rechenmacher}, {Quinn},
  {Richards}, {Richmond}, {Rivetta}, {Rockosi}, {Ruthmansdorfer}, {Sandford},
  {Schlegel}, {Schneider}, {Sekiguchi}, {Sergey}, {Shimasaku}, {Siegmund},
  {Smee}, {Smith}, {Snedden}, {Stone}, {Stoughton}, {Strauss}, {Stubbs},
  {SubbaRao}, {Szalay}, {Szapudi}, {Szokoly}, {Thakar}, {Tremonti}, {Tucker},
  {Uomoto}, {Vanden Berk}, {Vogeley}, {Waddell}, {Wang}, {Watanabe},
  {Weinberg}, {Yanny}, {Yasuda}, \& {SDSS Collaboration}}]{york00}
{York}, D.~G., et~al.\ 2000, \aj, 120, 1579

\bibitem[{{Zacharias} {et~al.}(2004){Zacharias}, {Urban}, {Zacharias},
  {Wycoff}, {Hall}, {Monet}, \& {Rafferty}}]{zacharias04}
{Zacharias}, N., {Urban}, S.~E., {Zacharias}, M.~I., {Wycoff}, G.~L., {Hall},
  D.~M., {Monet}, D.~G., \& {Rafferty}, T.~J. 2004, \aj, 127, 3043



\end{thebibliography}

\end{document}